\documentclass[AMA,STIX1COL]{WileyNJD-v2}

\articletype{Article Type}%

\received{1 January 2018}
\revised{xx xxxxx xxx}
\accepted{xx xxxxx xxx}
\usepackage{subfigure}
\usepackage{euscript}
\usepackage{tikz}
\usepackage{pgfplots}

\usepackage{float}

\definecolor{mblue}{rgb}{0,0.4470,0.7410}
\definecolor{morange}{rgb}{0.8500,0.3250,0.0980}
\definecolor{myellow}{rgb}{0.9290,0.6940,0.1250}
\definecolor{mpurple}{rgb}{0.4940,0.1840,0.5560}
\definecolor{mgreen}{rgb}{0.4660,0.6740,0.1880}
\definecolor{mcyan}{rgb}{0.3010,0.7450,0.9330}
\definecolor{mred}{rgb}{0.6350,0.0780,0.1840}
\definecolor{mgray}{rgb}{0.75,0.75,0.75}
\DeclareRobustCommand{\legendline}[1]{
\tikz[#1,line width=0.4mm,baseline=-0.5ex]{\draw (0,0) -- (.35,0);}
}

\newcommand{\R} {\mathbb{R}}

\newcommand{\X} {\mathbb{X}}
\newcommand{\Y} {\mathbb{Y}}

\newcommand{\U} {\mathbb{U}}

\newcommand{\bbma} {\begin{bmatrix} }
\newcommand{\ebma} {\end{bmatrix}}
\newcommand{\m}{\mathrm}
\newcommand{\mas}[2]{\left[\begin{array}{#1}#2\end{array}\right]}

\newcommand{\dimp}{n_\mathrm{p}}

\newcommand{\MOD}[1]{{#1}}  	% Changed
\begin{document}
\graphicspath{{figures/}}

\title{{LPV} Modeling of Nonlinear Systems: \\A Multi-Path Feedback Linearization Approach}

\author[1,6]{Hossam S. Abbas}

\author[2]{Roland T\'{o}th*}

\author[3]{Mih\'{a}ly Petreczky}

\author[4]{Nader Meskin}

\author[5]{Javad Mohammadpour Velni}

\author[2]{Patrick J.W.~Koelewijn}

\authormark{Hossam S. Abbas \textsc{et al}}

\address[1]{\orgdiv{Institute for Electrical Engineering in Medicine}, \orgname{University of
L{\"u}beck}, \orgaddress{\country{Germany}}}

\address[2]{\orgdiv{Control Systems Group, Department of Electrical Engineering}, \orgname{Eindhoven
University of Technology}, \orgaddress{\country{The Netherlands}}}

\address[3]{\orgdiv{Centre de Recherche en Informatique}, \orgname{Signal et Automatique de Lille (CRIStAL)}, \orgaddress{\country{France}}}

\address[4]{\orgdiv{Department of Electrical Engineering, College of Engineering}, \orgname{Qatar University}, \orgaddress{\country{Qatar}}}

\address[5]{\orgdiv{School of Electrical and Computer Engineering}, \orgname{The University of Georgia}, \orgaddress{\state{GA}, \country{USA}}}

\address[6]{\orgdiv{Electrical Engineering Department, Faculty
of Engineering}, \orgname{Assiut University}, \orgaddress{\country{Egypt}}}

\corres{*R. T\'{o}th, Control Systems Group, Department of Electrical Engineering, Eindhoven
University of Technology, P.O. Box 513, 5600 MB, Eindhoven, The
Netherlands. 
\email{r.toth@tue.nl}}

\fundingInfo{The first author H. S. Abbas is funded by the Deutsche Forschungsgemeinschaft
(DFG, German Research Foundation) under project No. 419290163.}

%\fundingInfo{NPRP grant (No. 5-574-2-233) from the Qatar National Research Fund (a member of the Qatar Foundation) and from the European Research Council (ERC) under the European Union's Horizon 2020 research and innovation programme (grant agreement No. 714663). The statements made herein are solely the responsibility of the authors.}

\abstract[Summary]{
This paper introduces a systematic approach to synthesize linear parameter-varying (LPV) representations of nonlinear (NL) systems which are described by input affine state-space (SS) representations. The conversion approach results in LPV-SS representations in the observable canonical form. Based on the relative degree concept, first the SS description of a given NL representation is transformed to a normal form.  In the SISO case, all nonlinearities of the original system are embedded into one NL function, which  is factorized, based on a proposed algorithm, to construct an LPV representation of the original NL system. The overall %transformation 
procedure yields an LPV model in which the scheduling variable depends on the inputs and outputs of the system and their derivatives, achieving a practically applicable transformation of the model \MOD{in case of low order derivatives}. In addition, if the states of the NL model can be measured or estimated, then a modified procedure is proposed to provide LPV models scheduled by these states. Examples are included to demonstrate both approaches.}


\maketitle

%\footnotetext{\textbf{Abbreviations:} LPV, linear parameter-varying; NL, nonlinear; SS, state-space}

%%%%%%%%%%%%%%%%%%%%%%%%%%%%%%%%%%%%%%%%%%%%%%%%%%%%%%%%%%%%%%%%%%%%%%%%%%%%%%%%

\section{Introduction}\label{sec:intro}

The \emph{linear parameter-varying} (LPV) framework was introduced to address the  control of  \emph{nonlinear} (NL) and \emph{time-varying} (TV) systems using the extensions of powerful \emph{linear
time-invariant} (LTI) approaches such as $\mathcal{H}_2/\mathcal{H}_\infty$ optimal control and model predictive control, see \emph{e.g.}, \cite{Carsten96,mohammadpour2011lpv,Besselmann2012,Toth17AUT,Toth18JRNP}. LPV systems are dynamical models capable of describing NL/TV behaviors in terms of a linear structure. Signal relations between the inputs and outputs in an LPV representation are assumed to be linear, but, at the same time, dependent on a so-called \emph{scheduling variable} $p$ ($n_\mathrm{p}$-dimensional signal), which is assumed to be measurable and free (external) in the modeled system and taking values from a so-called \emph{scheduling region} $\mathbb{P}\subseteq \mathbb{R}^{n_\mathrm{p}}$, often restricted to be a compact set. In this way, variation of $p$ represents time-variance, changing operating conditions, etc., and aims at the embedding  of the original NL/TV behavior into the solution set of an LPV system representation\cite{Toth2010SpringerBook,Rugh00}. While the former objective is pursued by the so-called \emph{global} LPV modeling approaches, alternatively, one can aim at the approximation of the NL/TV behavior by the interpolation of various linearizations of the system around operating points or signal trajectories, often referred to as \emph{local} modeling, see, \emph{e.g.}, \cite{Toth13JPC,Petersson2012,Shamma90c}.

\begin{figure*}
  \centerline{\subfigure[Original plant.]{\includegraphics[scale=0.45]{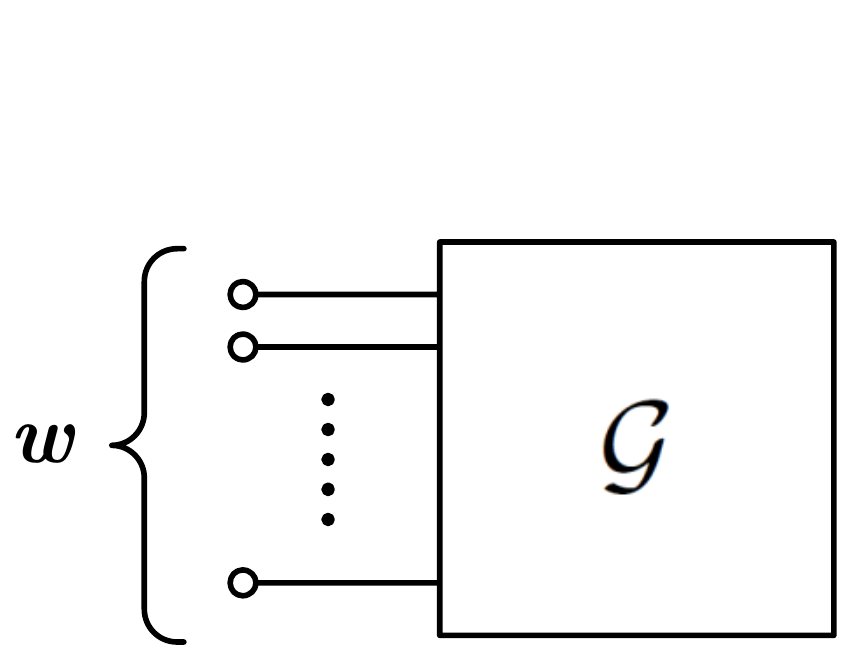}}\quad\quad \subfigure[Characterization of $p$.]{\includegraphics[trim=27.3cm 20.45cm 5.8cm 4.6cm, clip=true, scale=0.60]{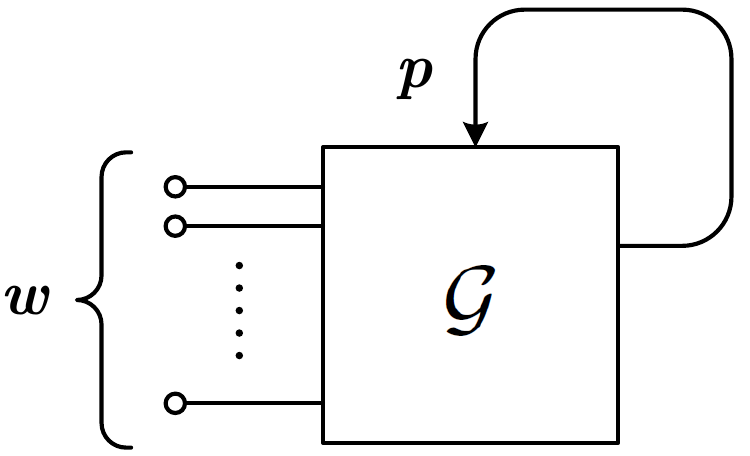}} \quad \subfigure[LPV form by disconnecting $p$.]{\includegraphics[trim=19.5cm 14.65cm 14.8cm 10cm, clip=true, scale=0.60]{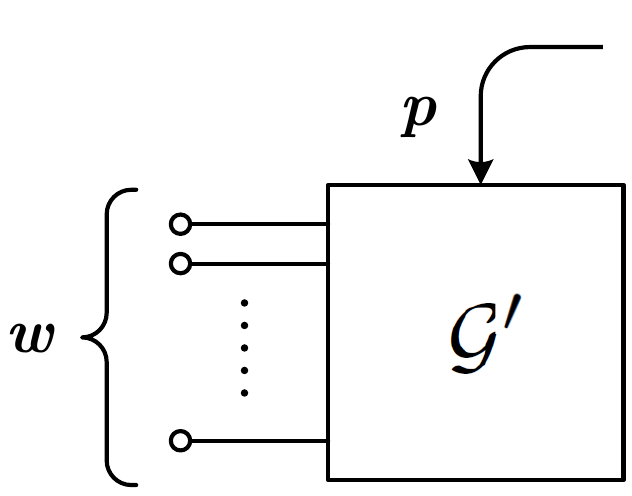}}\quad\quad \subfigure[Relation of the resulting behaviors.]{\includegraphics[scale=0.35]{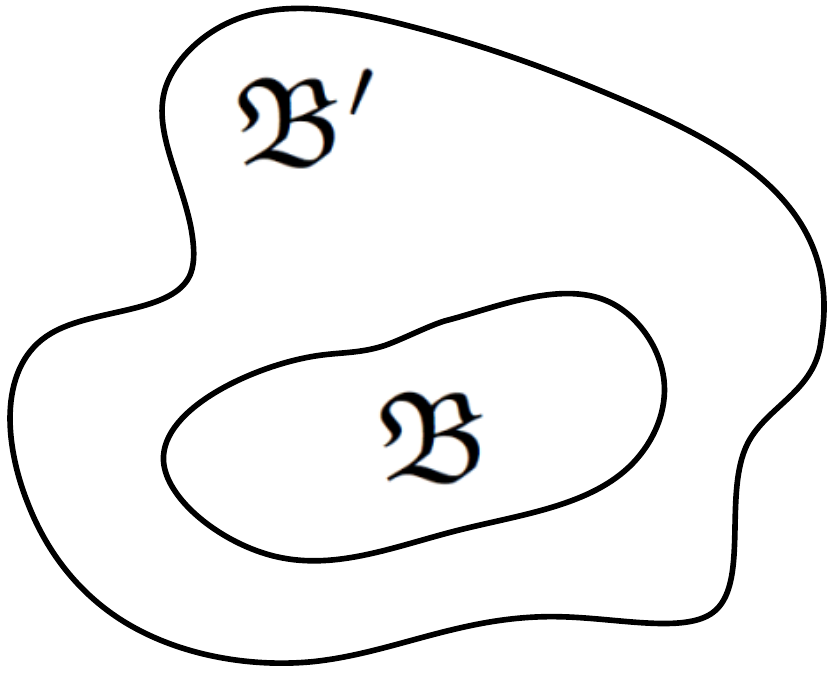}}  }
  \caption{The concept of LPV modeling.}\label{fig:LPVmod} 
\end{figure*}

\begin{figure}
\centerline{\includegraphics[scale=0.59]{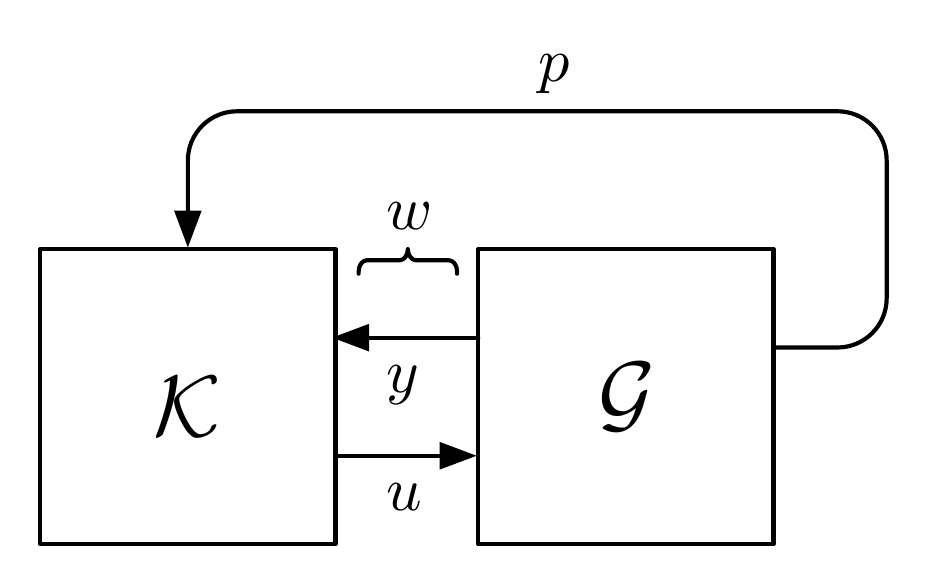}} 
  \caption{The concept of LPV control.}\label{fig:LPVcont} 
\end{figure}

For the global modeling methodology we intend to investigate in this paper, 
it is important to shed light on the often vaguely defined concept of LPV embedding. Assume that a continuous-time system $\mathcal{G}$, depicted in Fig.\ \ref{fig:LPVmod}.a,  is given which describes the (possibly nonlinear) dynamical relation between the signals $w:\R\rightarrow\mathbb{W}$, where $\mathbb{W}$ is a given set. For example consider the forced  \emph{Van der Pol} equation \cite{Bruzelius2004}:
\begin{subequations}
\label{eq:vdp}
\begin{align}
\dot{x}_1 &= x_2,\\
\dot{x}_2 &= -x_1+\alpha (1-x_1^2)x_2 + u,\\
y&=x_1,
\end{align}
\end{subequations} 
where, $\mas{cc}{x_1 & x_2}^\top:\mathbb{R}\rightarrow \mathbb{R}^2$ is the state variable, while $w=\mas{cc}{u  & y}^\top$ are the inputs and outputs of the system with $\mathbb{W}=\mathbb{R}^2$. Let  $\mathfrak{B}\subseteq\mathbb{W}^\R$ ($\mathbb{W}^\R$ stands for all maps from
$\R$ to $\mathbb{W}$) containing all trajectories of $w$ that are compatible with $\mathcal{G}$, \emph{i.e.}, they are solutions of \eqref{eq:vdp}.  
We call $\mathfrak{B}$ the (manifest) behavior of the system $\mathcal{G}$. A common practice in LPV modeling is to introduce an auxiliary variable $p$, with range $\mathbb{P}$, and reformulate $\mathcal{G}$ as shown in Fig.\ \ref{fig:LPVmod}.b, where it holds true that if the loop is disconnected and $p$ is assumed to be a known signal as in Fig.~\ref{fig:LPVmod}.c, then the ``remaining'' relations of $w$ are linear. This can be achieved in \eqref{eq:vdp} by taking, as a possible choice, $p=x_1=y$:
\begin{equation}
\label{eq:vdp2}
\mas{c}{\dot{x}\\ y} = \mas{cc|c}{0 & 1 & 0\\ -1 & \alpha(1-p^2) & 1 \\ \hline 1 & 0 & 0 } \mas{c}{x \\ u}. 
\end{equation}
Applying this reformulation with a disconnected $p$ and assuming that all trajectories of $p$ are allowed, \emph{i.e.}, $p$ is a free variable with $p\in\mathbb{P}^{\R}$ independent of $y$, the possible trajectories of this reformulated system $\mathcal{G}^\prime$ form a solution set of \eqref{eq:vdp2}, denoted as $\mathfrak{B}^\prime$, which  contains $\mathfrak{B}$ as visualized in Fig.\ \ref{fig:LPVmod}.d. This concept of formulating $\mathcal{G}^\prime$, a linear, but $p$-dependent description of $\mathcal{G}$, enables the use of simple stability analysis and convex controller synthesis, see \emph{e.g.}, \cite{Carsten96,mohammadpour2011lpv,Besselmann2012}, which can be conservative w.r.t.\ $\mathcal{G}$, but computationally more attractive and robust than other approaches directly addressing $\mathfrak{B}$. Control synthesis based on the above mentioned modeling procedure results in the implementation of an LPV controller  $\mathcal{K}$ visualized in Figure \ref{fig:LPVcont}.  It is obvious that 
a key assumption is that $p$ must be ``observable'' from the real system. The  observed value of $p$ is required to complete the hidden relation of $p$ to the other variables in \eqref{eq:vdp2}  and enable a linear controller to schedule its behavior according to $p$ to regulate \eqref{eq:vdp}. Hence, this can be seen as a multi-path feedback linearization, similar to the well-known approach in NL system theory, see \cite{Is95}, as the obtained information from the system in terms of $p$ is fed back to arrive to a varying linear relation \eqref{eq:vdp2} (in contrast with the NL theory where the resulting behavior is intended to be LTI).

Following the above procedure, the scheduling variable $p$ itself can appear in many different relations w.r.t.\ the original variables $w$. 
If $p$ is a free variable w.r.t.\ $\mathcal{G}$, e.g., wind speed for a wind turbine \cite{Bianchi07}, then we can speak about a \emph{true parameter-varying system} without conservativeness. However, in many practical applications, like in our example, it happens that $p$ depends on other signals, like inputs, outputs or states of the modeled system (\emph{e.g.}, operating conditions). Such situations are often warningly labeled to be \emph{quasi}-LPV (q-LPV). Based on the toy example \eqref{eq:vdp2}, what really happens in those cases 
is that the assumed freedom of $p$ only introduces conservativeness in the embedding of the nonlinear behavior. 
Hence, one important objective of LPV modeling, besides achieving complete embedding, is to \emph{minimize} such \emph{conservativeness}. Furthermore, it is often tempting to choose state variables as $p$ that are hardly measurable or cannot be reliably estimated from the measurements. For example, in \eqref{eq:vdp}, we could have chosen $p=x_1x_2$ which is not directly measurable. 
 Such choices can result in a loss of internal stability of the closed-loop system,  as an uncontrollable/unobservable mode can be introduced between the observer used to track $p$ and the controller that schedules based on it. These problems often undermine the results that can be obtained in practical applications of the LPV methodology leaving conversion of NL models to LPV representations to be a cumbersome procedure with many pitfalls for the regular user \cite{Rugh91,Toth2010SpringerBook}.

Existing approaches for \emph{global} LPV modeling of NL dynamical systems can be classified into two main categories: \emph{substitution based transformation} (SBT) methods  \cite{Shamma93, Papageorgiou00,Bokor2007,Toth11ACC_Philips,Rugh00,Leith98b,Marcos04} and \emph{automated conversion procedures} \cite{Donida2009,Kwiatkowski06,Toth2010SpringerBook,Hoffmann2015}.  For a detailed comparison, see \cite{Toth2010SpringerBook}.
In general\footnote{Except for the decision tree algorithm in \cite{Toth2010SpringerBook} and \cite{Hoffmann2015}.}, the existing techniques do not pay serious attention to several issues regarding the resulting LPV models, namely: 
how the scheduling variable and its bounds are chosen, what is the relation between these choices and the behavior of the system including the practical implementation of LPV controllers based on them, and the usefulness of the resulting LPV form for control synthesis or as a source of model structure information for identification. In addition, most techniques are based on ad-hoc mathematical manipulations (non-unique and non-systematic) and require a serious level of experience to be used.

In this paper\footnote{Preliminary ideas leading to the theorems presented in this paper appeared in the conference contribution \cite{Toth14IFAC}.}, inspired by the strong link between feedback linearization of NL representations \cite{Is95} and global LPV modeling, our objective is to provide systematic LPV embedding of the behavior of NL representations such that 
\begin{itemize}
\item the precise relationship between the behavior of the NL representation and the LPV representation is mathematically formalized;
\item the choice of $p$ and its bounds are explicit. 
\end{itemize}
Specifically, a systematic procedure is proposed to convert control affine NL-SS representations into state minimal LPV-SS representations in an observable canonical form. 
A particular advantage of this canonical form is that it can be directly converted into an equivalent LPV-IO form using the recently  developed LPV realization theory \cite{Toth2010SpringerBook} and hence it is highly useful for both LPV control synthesis (due to the SS form) and model structure selection in LPV identification (due to a direct LPV-IO conversion). The method is based on transforming the states of a given NL representation into a normal form such that, in the SISO case, all nonlinearities in the NL model are realized in only one NL term. Then, an exact substitution-based technique is presented to provide the LPV model. The state transformation leads to the systematic construction of scheduling signals.  More precisely, the scheduling signals depend either on the inputs, outputs, and their derivatives, or on  some of the observable states of the original NL representation. Explanation on why such a scheduling construction is practically useful will be provided in detail. Examples are also given to illustrate the procedure.

%<<<<<<<<<<<<<<<<<<<<<<<<<<<<<<<<<<<<>>>>>>>>>>>>>>>>>>>>>>>>>>>>>>>>>>>>>>>>>> 

\section{LPV representations}\label{s:lpvrep}
%<<<<<<<<<<<<<<<<<<<<<<<<<<<<<<<<<<<<>>>>>>>>>>>>>>>>>>>>>>>>>>>>>>>>>>>>>>>>>>
As the first step, we define the class of the considered LPV system representations and their associated solution sets, \emph{i.e.}, behaviors,  which will be used to describe/embed the solution set of nonlinear systems, further defined in Section  \ref{s:nl2lpvobvs_n}. 
%<<<<<<<<<<<<<<<<<<<<<<<<<<<<<<<<<<<<>>>>>>>>>>>>>>>>>>>>>>>>>>>>>>>>>>>>>>>>>>

\subsection{Mathematical preliminaries}\label{mathpre}
%<<<<<<<<<<<<<<<<<<<<<<<<<<<<<<<<<<<<>>>>>>>>>>>>>>>>>>>>>>>>>>>>>>>>>>>>>>>>>>

% Construction of the function set should be rewritten. It is not correct in this way!!!

Let $\mathcal{C}_k ({\R},\mathbb{W})$ be the space of $k$-times continuously differentiable real functions $w:\R\rightarrow\mathbb{W}\subseteq  {\R^{n_\mathrm{w}}}$ with left compact support that satisfy  $\frac{d^{i}}{dt^i} w(t) \in \mathbb{W} $ for all $t\in\mathbb{R}$ and $i\in\mathbb{I}_1^{k}=\{1,\ldots,k\}$.
Let $\mathbb{P}$ be an open subset of $\mathbb{R}^{\dimp}$ and let $\mathcal{R}_k(\mathbb{P})$ denote the \emph{set} of real-analytic functions of the form 
$f:\mathbb{P}^k \rightarrow \mathbb{R}$ in $\dimp k$ variables. For  $\hat{k}> k$, any  $f \in \mathcal{R}_k(\mathbb{P})$ is called equivalent with a $\hat{f} \in \mathcal{R}_{\hat{k}}(\mathbb{P})$ if
$\hat{f}(\eta_1,\ldots,\eta_{\hat{k}})=f(\eta_1,\ldots,\eta_k)$ for all $\eta_1,\ldots,\eta_{k}\in\mathbb{P}$, as $\hat{f}$ is not  \emph{essentially dependent} on its arguments. Define the set operator $\circleddash$, such that $\mathcal{R}_{k+1}(\mathbb{P}) \circleddash  \mathcal{R}_{k}(\mathbb{P}) $ contains all $f\in \mathcal{R}_{k+1}(\mathbb{P}) $ not equivalent with any element of   $\mathcal{R}_{k}(\mathbb{P}) $.
This prompts to considering the set $\mathcal{R}(\mathbb{P})= \bigcup_{k=0}^{\infty} \mathcal{R}_k(\mathbb{P}) \circleddash  \mathcal{R}_{k-1}(\mathbb{P})$ where $\mathcal{R}_{0}(\mathbb{P})=\mathbb{R}$ and $ \mathcal{R}_{-1}(\mathbb{P})=\emptyset$. We can define addition and multiplication in $\mathcal{R}(\mathbb{P})$ analogous to that of \cite{Toth11_LPVBehav}:
if $f_1, f_2 \in \mathcal{R}(\mathbb{P})$, then $f_i \in \mathcal{R}_{k_i} (\mathbb{P}) \circleddash   \mathcal{R}_{k_i-1}(\mathbb{P})$, for some integer $k_i \geq 0$, $i=1,2$, and, by taking $k=\max\{k_1,k_2\}$, the equivalence described above implies that there exist equivalent representations of these functions in 
$ \mathcal{R}_k(\mathbb{P})$. Then $f_1+f_2, f_1 \cdot f_2$ can be defined as the usual addition and multiplication of functions in $ \mathcal{R}_k(\mathbb{P})$ and the result, in terms of the equivalence, is considered to be a $f\in  \mathcal{R}(\mathbb{P})$. 
For a  $p \in \mathcal{C}_{\infty}({\R},\mathbb{P})$, we define the following notation: if $f \in \mathcal{R}(\mathbb{P})$, then
$f \diamond p: \mathbb{R} \rightarrow \mathbb{R}$ is 
\begin{equation}
\forall t \in \mathbb{R}:\quad  (f \diamond p)(t) =f\left(p(t),\tfrac{d}{dt}p(t),\ldots, \tfrac{d^k}{dt^k} p(t) \right),
\end{equation}
where $k$ is an integer such  that $f \in \mathcal{R}_{k}(\mathbb{P})\ \MOD{\circleddash}\ \mathcal{R}_{k-1}(\mathbb{P})$. 
We denote by $\mathcal{R}^{k \times l}(\mathbb{P})$ the set of all $k \times l$ matrices whose entries are elements of $\mathcal{R}(\mathbb{P})$ which also extends 
the operator $\diamond$ to matrices whose entries are functions from $\mathcal{R}(\mathbb{P})$.

%<<<<<<<<<<<<<<<<<<<<<<<<<<<<<<<<<<<<<<<<<<<<<<<<<<<<<< 
\subsection{State-space representation}\label{ss:lpvssrep}
%<<<<<<<<<<<<<<<<<<<<<<<<<<<<<<<<<<<<>>>>>>>>>>>>>>>>>>>>>>>>>>>>>>>>>>>>>>>>>>
For the sake of simplicity for defining the embedding of the dynamics of an NL system into the solution set of an LPV representation, we will introduce a slightly extended definition of LPV state-space representations compared to the regular definitions treated in the literature \cite{Rugh00,Shamma90c}. 
\begin{definition}[LPV-SS representation] \label{def:CT_SS_LPV_rep} A continuous-time LPV-SS representation 
with an open scheduling region $\mathbb{P}$ of dimension $\dimp$ is a tuple of matrices of analytic functions:
\begin{equation} \label{eq:dep}
\mas{c|c}{\EuScript{A} & \EuScript{B}\\ \hline \EuScript{C} & \EuScript{D}}  \in \mas{c|c}{\mathcal{R}^{n_\mathrm{z}\times n_\mathrm{z}}(\mathbb{P}) & \mathcal{R}^{n_\mathrm{z}\times n_\mathrm{u}}(\mathbb{P}) \\ \hline \mathcal{R}^{n_\mathrm{y}\times n_\mathrm{z}}(\mathbb{P}) & \mathcal{R}^{n_\mathrm{y}\times n_\mathrm{u}}(\mathbb{P})}.
\end{equation} 
A solution of this representation is a tuple $(u,z,y,p)\in  \mathcal{C}_{n_\mathrm{z}}(\R,\U\times\mathbb{Z} \times\Y) \times  \mathcal{C}_\infty(\R,\mathbb{P})$ such that
\begin{subequations}
\label{eq:ch3:04all}
\begin{eqnarray}
\frac{d}{d t}z &=&(\EuScript{A}\diamond p) z +(\EuScript{B}\diamond p) u ,  \label{eq:ch3:03} \\
y &=&(\EuScript{C} \diamond p) z +(\EuScript{D} \diamond p ) u,
\label{eq:ch3:04}
\end{eqnarray}
\end{subequations}
where $z$ is the state vector\footnote{We use $z$ to denote the state vector in an LPV-SS representation. This allows  later to distinguish $z$ from the state vector $x$ associated with an NL-SS representation.},
$\mathbb{Z}=\R^{n_\mathrm{z}}$ is the state space, $u:\mathbb{R}\rightarrow \U=\mathbb{R}^{n_\mathrm{u}}$ is the input while $y:\mathbb{R}\rightarrow \Y=\mathbb{R}^{n_\mathrm{y}}$ is the output of the represented system.
We denote by 
\begin{equation}
\mathfrak{B}_\mathrm{SS}=\left\{ (u,z,y,p)\in  \mathcal{C}_{n_\mathrm{z}}(\R,\U\times\mathbb{Z} \times\Y) \times  \mathcal{C}_\infty(\R,\mathbb{P}) \mid 
\text{\eqref{eq:ch3:03}-\eqref{eq:ch3:04} hold} \right\},
\end{equation} 
the solution set (latent behavior) of  \eqref{eq:ch3:03}-\eqref{eq:ch3:04}. 
\end{definition}
Note that in the above defined SS representation, the operator $\diamond$ 
 expresses the dependence of the state-space matrix functions along a scheduling trajectory $p$ and its derivatives; in other words, it expresses a dynamic mapping between $p$ and $(\EuScript{A},\EuScript{B},\EuScript{C},\EuScript{D})$. We refer to this dynamic mapping between the scheduling signal and the system matrices as \emph{dynamic dependence}, whereas the dependence on the value of $p(t)$ only is referred to as \emph{static dependence}. The latter is used in the conventional definitions that can be found in the literature \cite{Rugh00,Shamma90c}, however, we need the notion of dynamic dependence here to show how systematic embedding of NL systems can be achieved by LPV models. Moreover, LPV models with dynamic dependence arise naturally as a result of system manipulations, such as state transformations, observability, controllability canonical forms, etc. \cite{Toth11_LPVBehav}.
For technical reasons, in this paper we work with LPV-SS representations in observable canonical form. As its name suggests, an LPV model in observable 
canonical form is state observable and it  allows a simple conversion to \emph{input-output} (IO) representations. The latter is important for system identification,
since IO representations are easier to identify than state-space models. Conditions for existence of a state-space isomorphism transforming an LPV-SS representation to an observable canonical form are
discussed in \cite{Toth10_SSReal,Toth11_LPVBehav}.
The matrices, associated with the observability canonical representation of \eqref{eq:ch3:04all} in the SISO case, under the assumption of minimality of \eqref{eq:ch3:04all}, are given by \cite{Toth2010SpringerBook}: 
\begin{equation}\label{e:lpv_obvs}
\mas{c|c}{\EuScript{A}&\EuScript{B}\\ \hline \EuScript{C} & \EuScript{D}}=
\mas{cccc|c}{
0 & 1 & \dots &  0 & \beta_{n_\mathrm{z}-1}\\
\vdots & \vdots & \ddots & \vdots & \vdots    \\
0 & 0 & \dots & 1 & \beta_{1}   \\ 
\alpha_0 & \alpha_1 &  \dots &  \alpha_{n_\mathrm{z}-1} & \beta_{0}  \\ \hline
1 & 0 & \dots & 0 & \beta_{n_\mathrm{z}} }
,
\end{equation}
where $\{\alpha_i\}_{i=0}^{n_\mathrm{z}-1}$ and  $\{\beta_j\}_{j=0}^{n_\mathrm{z}-1}$ are analytic functions in $\mathcal{R}(\mathbb{P})$. A special case of \eqref{e:lpv_obvs}, when $\beta_{n_\mathrm{z}}=\ldots =\beta_{1}=0$, is given by
\begin{equation}\label{e:lpv_obvs_n}
\mas{c|c}{\EuScript{A}&\EuScript{B}\\ \hline \EuScript{C} & \EuScript{D}} =
\mas{cccc|c}{
0 & 1 & \dots &  0 & 0 \\
\vdots & \vdots & \ddots & \vdots & \vdots \\
0 & 0 & \dots & 1 & 0  \\
\alpha_0\ & \alpha_1 &  \dots &  \alpha_{n_\mathrm{z}-1} & \beta_{0} \\ \hline
1 & 0 & \dots & 0 & 0 
},
\end{equation}
which is of particular importance in this work as demonstrated later. In the sequel, we refer to the forms \eqref{e:lpv_obvs} and \eqref{e:lpv_obvs_n} as the \emph{full and simplified observability forms}, respectively. In this paper, we present a method for transforming a nonlinear system to LPV simplified observability form and another method which yields an LPV representation
in full observability form.
%<<<<<<<<<<<<<<<<<<<<<<<<<<<<<<<<<<<<>>>>>>>>>>>>>>>>>>>>>>>>>>>>>>>>>>>>>>>>>>
\section{Conversion to the simplified observability form} \label{s:nl2lpvobvs_n}
%<<<<<<<<<<<<<<<<<<<<<<<<<<<<<<<<<<<<>>>>>>>>>>>>>>>>>>>>>>>>>>>>>>>>>>>>>>>>>>
In this section, we discuss conversion of input-affine nonlinear models to simplified LPV observability canonical forms.
\subsection{The problem setting}
Consider a SISO NL system $\mathcal{G}$ represented in the form of  
\begin{subequations}\label{e:nl}
\begin{align}
  \frac{d}{dt} x &=f(x)+g(x)u, \label{e:nl1}\\
    y &=h(x), \label{e:nl2}
    \end{align}
\end{subequations} 
where $f,g:\mathbb{X}\rightarrow \R^{n_\mathrm{x}}$ and  $h:\mathbb{X} \rightarrow \R$ are real analytic functions, $\mathbb{X}$ is an open subset of $\R^{n_\mathrm{x}}$ and $u:\mathbb{R}\rightarrow \U \subseteq\mathbb{R}$ is the input with  $y:\mathbb{R}\rightarrow \Y \subseteq \mathbb{R}$ being the output signal and $x:\mathbb{R}\rightarrow \mathbb{X}$ is the state variable.  We consider the solutions of  \eqref{e:nl} in the following sense 
\begin{equation}
\mathfrak{C}_\mathrm{SS}=\left\{ (u,x,y)\in  \mathcal{C}_{n_\mathrm{x}}(\R,\U\times\mathbb{X} \times\Y) \mid
\text{(\ref{e:nl1}--b) hold
for all $t\in\mathbb{R}$} \right\}.
\end{equation} 
The form \eqref{e:nl} represents a rather general class of NL systems, commonly referred to as input-affine systems, which includes common models of mechanical systems \cite{Nijmeijer96} and many first-principles models in process control \cite{HeSe98}. More general representation of NL systems characterized by $\dot{x} =f(x,u)$, with $f: \mathbb{X}\times\U\rightarrow \R^{n_\mathrm{x}}$ being an analytic vector field, can be rewritten in the input affine form  \eqref{e:nl} according to the procedure detailed in  \cite{Nijmeijer96}.
Furthermore, in \eqref{e:nl2}, there is no direct feedthrough term as, w.l.o.g., such feedthrough terms can be easily eliminated via the projection of $y$.

To achieve our objective, \emph{i.e.}, to embed the dynamical behavior of NL systems represented by \eqref{e:nl} into the solution set of an LPV-SS representation in a simplified  observable canonical form  given by  \eqref{e:lpv_obvs_n}, we intend to use the concept of the embedding principle discussed in Section \ref{sec:intro} to develop multi-path feedback linearization of \eqref{e:nl}.  Before going into the mathematical details, we present the main idea informally.
Consider a solution $(x,y,u)$ of \eqref{e:nl}, and define 
\begin{subequations}
\label{eq:zv}
\begin{align} z&=\left[\begin{array}{cccc} y & \frac{d}{dt} y & \ldots & \frac{d^{n_{\mathrm x}}}{dt^{n_{\mathrm x}}} y \end{array}\right]^\top,\\
v&=\left[\begin{array}{cccc} u & \frac{d}{dt} u & \ldots & \frac{d^{n_{\mathrm x}}}{dt^{n_{\mathrm x}}} u \end{array}\right]^\top.
\end{align}
\end{subequations}
Let \eqref{e:nl} be observable, \emph{i.e.},  $x=\Psi(z,v)$
for some map $\Psi$ and let $\Phi$ (an implicit function of $\Psi$) be such that  
\begin{equation}\label{e:phi}
    z = \Phi(x,v).
\end{equation}
Then, we can obtain a new state-space description of \eqref{e:nl}:
\begin{subequations}
\label{e:nl_new}
\begin{align}
        \frac{d}{dt}z_1  &= z_2, && \ldots & \frac{d}{dt}z_{n_\mathrm{x}-1}  &= z_{n_\mathrm{x}}, \\
&&        \frac{d}{dt}z_{n_\mathrm{x}} &= \bar{\alpha}(z, v), \label{eq:mfrt}  \\
      &&  y &= z_1 ,
 \end{align}
 \end{subequations}
 where $\bar{\alpha}$ is an analytical function,
such that if $(u,x,y)$ is a solution of \eqref{e:nl}, then 
$(u,z,y)$ is a solution of \eqref{e:nl_new} with $z$ and $x$ related by \eqref{e:phi}.
If $\bar\alpha$ can be factorized as
\begin{equation}
\bar{\alpha}(z, v)=\beta_0(z, v)u +
  \sum_{i=0}^{n_{\mathrm x}-1} \alpha_i(z, v)z_{i+1},
\end{equation}
for some analytic functions $\beta_0$ and $\{\alpha_i\}_{i=0}^{n_{\mathrm x}-1}$, then by setting $p=[\begin{array}{cc} y & u \end{array}]^\top$, and changing the ordering of the arguments of $\beta_0$ and $\{\alpha_i\}_{i=0}^{n_{\mathrm x}-1}$, \eqref{eq:mfrt} can be written as
\begin{equation}
   \frac{d}{dt} z_{n_{\mathrm x}}=\sum_{i=0}^{n_{\mathrm x}-1} (\alpha_i \diamond p)z_{i+1} + (\beta_0 \diamond p)u,
\end{equation}
which implies that $(u,z,y,p)$, with $z$ being related to $x$ by \eqref{e:phi} and $p=[\begin{array}{cc} y & u \end{array}]^\top$, is a solution
of an LPV observable canonical form \eqref{e:lpv_obvs_n} with $n_\mathrm{x}=n_\mathrm{z}$. As $p$ of the resulting LPV-SS model is composed of the output and input signals of the system, it is measurable/available in most real-world applications, \emph{i.e.}, the transformation yields an LPV-SS form that opens the possibility to design LPV controllers \MOD{for which implementation can avoid or mitigate the need for state measurements or scheduling observers}.

\subsection{Mathematical details of the construction }\label{sec:compem}

Below we present the ideas outlined above in a more rigorous way. First of all, note that we need to choose a point $x_0\in \mathbb{X}$ around which the embedding can be developed and its validity can be analyzed.  
From the point of view of controller synthesis, it is often desirable to consider $x_0=0$ so that any stabilizing
controller designed for the resulting LPV-SS form will aim at keeping the state of the original system in a neighborhood of $x_0$. To this end, we will make the following assumption. 

\begin{assumption}[Centering]
To simplify the discussion,  in the sequel, we will assume w.l.o.g.\ that $f(x_0)=0$ and $h(x_0)=0$.
\end{assumption}
Note that $f(x_0)=0$ can easily be achieved by state and input transformation, while $h(x_0)=0$ requires transformation of the output signal $y$.

\begin{definition}[$(\mathbb{U}_0,\mathbb{X}_0,\mathbb{Y}_0)$-admissible solutions]
Let  $\mathbb{X}_0$ be an open neighborhood of $x_0$ in $\mathbb{X}$.
Furthermore, choose open sets $0 \in \U_0\subseteq \R$, $0 \in \Y_0\subseteq \R$. 
A solution $(u,x,y)\in \mathfrak{C}_\mathrm{SS}$ of \eqref{e:nl} is said to be $(\mathbb{U}_0,\mathbb{X}_0,\mathbb{Y}_0)$-admissible, if
$u \in \mathcal{C}_{n_\mathrm{x}}(\R,\U_0)$, $x \in \mathcal{C}_\infty(\R,\mathbb{X}_0)$ and $y \in \mathcal{C}_{n_\mathrm{x}}(\R,\Y_0)$.
\end{definition}
Next, we recall from \cite{TeelPralyUniformObs,GauthierBornardUniformObs} the  notion of local uniform observability. 
\begin{definition}[Local uniform observability]
\label{e:nl:obs}
The representation \eqref{e:nl} is called locally uniformly observable on the open sets $x_0 \in \mathbb{X}_0 \subseteq \R^{n_\mathrm{x}}$, $0\in \U_0  \subseteq \R$, $0\in \Y_0  \subseteq \R$, if there exists
an analytic map \begin{equation} \Psi: (\Y_0 \times \U_0)^{n_\mathrm{x}} \rightarrow \X_0, \end{equation}
 such that for any $(\mathbb{U}_0,\mathbb{X}_0,\mathbb{Y}_0)$-admissible solution $(u,x,y)$ of \eqref{e:nl}, it holds that
 \begin{equation}
 \label{e:nl:obs:eq1}
  x=\Psi\left(\begin{bmatrix} y \\ u \end{bmatrix} , \frac{d}{dt} \begin{bmatrix} y \\ u \end{bmatrix}, \ldots, \frac{d^{n_\mathrm{x}-1}}{dt^{n_\mathrm{x}-1}} \begin{bmatrix} y \\ u \end{bmatrix}\right). 
  \end{equation}
 We will call the map $\Psi$ the \emph{$(\U_0,\X_0,\Y_0)$-observability map} or \emph{observability map}, if $(\U_0,\X_0,\Y_0)$ is clear from the context and call \eqref{e:nl} \emph{locally uniformly observable}, if it is locally uniformly observable on $(\U_0,\X_0,\Y_0)$ for some open sets $\U_0,\X_0,\Y_0$.
\end{definition}
 If \eqref{e:nl} is locally uniformly observable, then it is possible to express the $n_\mathrm{x}$-th derivative of its output
 $y$ as a function of  $\{\frac{d^i}{dt^i} y\}_{i=0}^{n_\mathrm{x}-1}$  and $\{\frac{d^j}{dt^j} u\}_{j=0}^{n_\mathrm{x}-1}$. In order to present the construction
 formally, we define the following collection of functions.
 \begin{definition}[Output derivative function]
\label{e:nl:out:der}
  For each $k\in\mathbb{N}$, define the functions $\Phi_k:\mathbb{X} \times \U^k \rightarrow \Y$ as follows:
  \begin{subequations}
  \begin{align}
  \Phi_0(x) =\ &h(x), \\
  \Phi_{k}(x,v_1,\ldots,v_{k})  =
\ &\!\!\sum_{i=1}^{n_\mathrm{x}}\! \left[ \left( f_i(x)+g_i(x)v_1 \right)\frac{\partial  \Phi_{k-1}}{\partial x_i}
  (x,v_1,\ldots,v_{k-1}) + \sum_{j=1}^{k-1} v_{j+1} \frac{\partial  \Phi_{k-1}}{\partial v_j} (x,v_1,\ldots,v_{k-1}) \right],
  \end{align}
  \end{subequations}
  where $f_i$ and $g_i$ denote the $i$-th element of these functions.  The map $\Phi_k$ will be called the $k$-th output derivative map.
 \end{definition}
For any $(\mathbb{U}_0,\mathbb{X}_0,\mathbb{Y}_0)$-admissible solution $(u,x,y)$ of \eqref{e:nl}:
  \begin{equation} \frac{d^k}{dt^k} y = \Phi_k(x,u,\tfrac{d}{dt} u, \ldots, \tfrac{d^{k-1}}{dt^{k-1}} u), \end{equation}
which leads to the following corollary: 
 \begin{corollary}[NL-IO realization]
 \label{e:nl:obs:cor}
  If \eqref{e:nl} is locally uniformly observable on $(\mathbb{U}_0,\mathbb{X}_0,\mathbb{Y}_0)$ with the observability function $\Psi$, 
  then for any $(\mathbb{U}_0,\mathbb{X}_0,\mathbb{Y}_0)$-admissible solution $(u,x,y)$ of \eqref{e:nl}:
  \begin{equation} \label{e:nl:obs:cor:eq1}
    \frac{d^{n_\mathrm{x}}}{dt^{n_\mathrm{x}}} y = \Gamma_{n_{\mathrm x}}\left(\begin{bmatrix} y \\ u \end{bmatrix} , \frac{d}{dt} \begin{bmatrix} y \\ u \end{bmatrix}, \ldots, \frac{d^{n_\mathrm{x}-1}}{dt^{n_\mathrm{x}-1}} \begin{bmatrix} y \\ u \end{bmatrix}\right) ,
 \end{equation}
  where the analytic map $\Gamma_{n_\mathrm{x}}:(\Y_0\times \U_0)^{n_{\mathrm x}} \rightarrow \Y_0$ is defined by 
  \begin{equation} 
    \Gamma_{n_{\mathrm x}}\left(\begin{bmatrix} \mu_1 \\ \upsilon_1 \end{bmatrix}, \ldots, \begin{bmatrix} \mu_{n_{\mathrm x}} \\ \upsilon_{n_{\mathrm x}} \end{bmatrix}\right) =   
     \Phi_{n_\mathrm{x}}\left( \Psi\left(\begin{bmatrix} \mu_1 \\ \upsilon_1 \end{bmatrix}, \ldots,  \begin{bmatrix} \mu_{n_{\mathrm x}} \\ \upsilon_{n_{\mathrm x}} \end{bmatrix} \right), \upsilon_1,\ldots,\upsilon_{n_{\mathrm x}} \right),
  \end{equation}
 for all $\mu_1,\ldots,\mu_{n_{\mathrm x}} \in \Y_0$ and $\upsilon_1,\ldots,\upsilon_{n_{\mathrm x}} \in \U_0$. 
 \end{corollary} 
 Corollary \ref{e:nl:obs:cor} paves the way to represent  $(\mathbb{U}_0,\mathbb{X}_0,\mathbb{Y}_0)$-admissible solutions of \eqref{e:nl} as solutions
 of an LPV observer canonical form. In order to present the precise result, we have to introduce some concepts related to factorization of functions.

Note that for a given open set $\mathbb{V}\subseteq \mathbb{R}^n$, any analytic function $f:\mathbb{V} \rightarrow \mathbb{R}$ can be decomposed as 
\begin{equation} \label{eq:ratform}
f(\xi)=\frac{N(\xi,\phi_1(\xi),\ldots,\phi_\tau(\xi))}{D(\xi,\phi_1(\xi),\ldots,\phi_\tau(\xi))}, \quad \forall \xi\in\mathbb{V},
\end{equation}
where $\xi$ is the indeterminate of $f$, $N$ and $D$ are polynomial maps: $\mathbb{R}^{n+\tau}\rightarrow \mathbb{R}$ and $\{\phi_i:\mathbb{V} \rightarrow \mathbb{R}\}_{i=1}^{\tau}$ are analytic functions. If \eqref{eq:ratform} holds, we will say that \emph{$f$ is rational w.r.t. $\{\phi_i\}_{i=1}^{\tau}$}. 
Note that if the functions $\{\phi_i\}_{i=1}^{\tau}$ are algebraically independent and $f$ is rational w.r.t.~to $\{\phi_i\}_{i=1}^{\tau}$, 
then there is a unique pair of co-prime polynomials $(N,D)$ which satisfies \eqref{eq:ratform}.

\begin{definition}[Factorization]\label{def:factor}
Consider a given open set $\mathbb{V}\subseteq \mathbb{R}^n$ and an analytic function $f:\mathbb{V} \rightarrow \mathbb{R}$, rational w.r.t.\ some analytic $\{\phi_i\}_{i=1}^{\tau}$ in terms of \eqref{eq:ratform}. 
Under $\{\phi_i\}_{i=1}^{\tau}$, factorization of $f$ with respect to the first $m$ variables is a tuple 
$(\{r_i:\mathbb{V} \rightarrow \mathbb{R} \}_{i=1}^{m},s:\mathbb{V} \rightarrow \mathbb{R})$ of analytic functions such that $r_i=M_i/D$ and $s=S/D$ in terms of \eqref{eq:ratform} with 
$\{M_{i}\}_{i=1}^{m}$, $D$ and $S$ being polynomials in $n+\tau$ variables $X_1, \ldots, X_{n+\tau}$ such that
\begin{equation} 
\label{def:factor:eq1}
   N=M_{1}X_1+\cdots + M_{m}X_m+S, 
\end{equation}
and, for all $i\in\mathbb{I}_1^{m}$, $M_i$ does not depend on $\{X_l\}_{l=i+1}^m$ and $S$ does not depend on $\{X_l\}_{l=1}^m$. 
\end{definition} 

The polynomials $\{M_i\}_{i=1}^m$ are the result of the division of $N$ by $\{X_l\}_{l=1}^m$ and
$S$ is the remainder of this division,  in the sense of \cite[Theorem 3, pp. 61-62]{Cox:2007}. As $\{X_l\}_{l=1}^m$ are monomials, a simplified form of the algorithm  described in \cite{Cox:2007} is available to compute the factorization (see Algorithm \ref{alg:fact1} later). Note that if $f$ is rational with respect to $\{\phi_i\}_{i=1}^{\tau}$, then  a factorization $(\{r_i\}_{i=1}^{m},s)$ with respect to the first $m$ variables always exists in the form of  \( f(\xi)=\sum_{i=1}^{m} r_i(\xi)\xi_i+s(\xi) \). This factorization depends on $\{\phi_i\}_{i=1}^{\tau}$, \emph{i.e.}, different choices of these functions will lead to different factorizations, the consequences of which will be discussed in Section \ref{sec:schechoice}. 

Introduce the selection matrix\footnote{A selection matrix contains zeros and a single element 1 in each row.} $\EuScript{R}\in\mathbb{R}^{2n_\mathrm{x}\times 2n_\mathrm{x}}$, which rearranges the arguments of $\Gamma_{n_\mathrm{x}}(\zeta)\!:\!(\Y_0\times \U_0)^{n_{\mathrm x}} \rightarrow \Y_0$ such that $\Gamma_{n_\mathrm{x}}(\EuScript{R}\xi)\!:\! \Y_0^{n_\mathrm{x}}\times \U_0^{n_\mathrm {x}}\rightarrow \mathbb{R}$ is equivalent with $\Gamma_{n_\mathrm{x}}$. Formally this means that for $\mu_1,\ldots,\mu_{n_{\mathrm x}} \in \Y_0$ and $\upsilon_1,\ldots,\upsilon_{n_{\mathrm x}} \in \U_0$, $\zeta=\begin{bmatrix} \mu_1\! &\! \upsilon_1\! &\! \ldots\! &\! \mu_{n_\mathrm{x}}\! &\! \upsilon_{n_\mathrm{x}}\end{bmatrix}=\EuScript{R}\xi$ where $\xi=\begin{bmatrix} \mu_1\! &\!  \ldots\! &\! \mu_{n_\mathrm{x}}\! &\! \upsilon_1\! & \! \ldots \! &\! \upsilon_{n_{\mathrm x}} \end{bmatrix}$. We identify the resulting function as $\Gamma_{n_\mathrm{x}}\! \circ\  \EuScript{R}$.
Furthermore, consider a set of functions $\{f_i:\mathbb{W}^l \rightarrow \mathbb{R}\}_{i=1}^\tau$, where $\mathbb{W}\subseteq \R^n$ is not necessarily open. 
The matrix $\EuScript{T} \in \mathbb{R}^{m \times n}$, $m \le n$, is called the \emph{selection matrix of the essential support of $\{f_i\}_{i=1}^\tau$ under} $\mathbb{W}$,
 if $\EuScript{T}$ has full row rank, and the functions $\{f_i(\zeta_1,\ldots,\zeta_l)\}_{i=1}^\tau$ with $\zeta_j\in\mathbb{W}$ depend only\footnote{$\forall \{\zeta_j^{(1)},\zeta_j^{(2)} \in \mathbb{W}\}_{j=1}^l$ and $\forall i\in\mathbb{I}_{1}^\tau$, $\zeta_j^{(1)}-\zeta_j^{(2)} \in \ker \EuScript{T}$ for all $j\in\mathbb{I}_{1}^{l} \implies f_i(\zeta_1^{(1)},\ldots,\zeta_l^{(1)})=f_i(\zeta_1^{(2)},\ldots,\zeta_l^{(2)})$ for all $i\in\mathbb{I}_1^{\tau}$} on $\EuScript{T}\zeta_j$. 
For example, if $f:\mathbb{R}^4\rightarrow \mathbb{R}$ depends only on its first and third arguments, then $\EuScript{T}={\tiny \mas{cccc}{1 & 0 & 0 &0\\ 0 & 0 &1&0}}$ is a 
selection matrix of the essential support of $f$ under $\mathbb{R}^4$, while $\EuScript{T}={\tiny \mas{cc}{1 & 0 }}$ is the selection matrix under $\mathbb{R}^2$. 
If $\EuScript{T}$ is a selection matrix for the essential support for $\{f_i\}_{i=1}^{\tau}$, then $\EuScript{T}^{-1}=\EuScript{T}^\top$ is  a selection matrix such that  $\EuScript{T}\cdot \EuScript{T}^{-1}=I$ and
we can identify the functions $\{f_i\}_{i=1}^{\tau}$ with the functions $\{f_i\ \circ\ \EuScript{T}^{-1}\}_{i=1}^{\tau}$. Note that while the former
are functions of $n\cdot l$ variables, the latter have $m\cdot l \le n \cdot l$ variables.  

\begin{theorem}[LPV embedding, simp. observability form]\label{thm:1}
Assume that \eqref{e:nl} is locally uniformly observable on $(\mathbb{U}_0,\mathbb{X}_0,\mathbb{Y}_0)$ with observability function $\Psi$. Furthermore, assume that there exists a set of analytic functions $\{\phi_i: \Y_0^{n_{\mathrm x}} \times \U_0^{n_{\mathrm x}} \rightarrow \mathbb{R} \}_{i=1}^{\tau}$ such that the map
$\Gamma_{n_\mathrm{x}} \circ\ \EuScript{R}$ in \eqref{e:nl:obs:cor:eq1}  is rational with respect to $\{\phi_i\}_{i=1}^{\tau}$.
 Let $(\{r_i\}_{i=1}^{n_\mathrm{x}+1},s)$ be a factorization of $\Gamma_{n_\mathrm{x}}\circ\ \EuScript{R}$ 
with respect to the first $n_\mathrm{x}+1$ variables. 
If $s=0$, \emph{i.e.}, factorization is possible without a remainder and $\EuScript{T}$ is the essential support of $\{r_i\circ \EuScript{R}^{-1}\}_{i=1}^{n_\mathrm{x}+1}$ under $\Y_0 \times \U_0$, then the LPV-SS representation \eqref{e:lpv_obvs_n} 
with \begin{subequations}
\begin{equation}
p=\EuScript{T} [\begin{array}{cc}y^\top & u^\top\end{array}]^\top  , \label{eq:sche:choice1} \vspace{-3mm}
 \end{equation} 
\begin{equation} \{\alpha_{i}:=r_{i+1}\ \circ\ \EuScript{R}^{-1}\ \circ\  \EuScript{T}^{-1} \}_{i=0}^{n_\mathrm{x}-1}, \ \ \ \ \beta_0 := r_{n_\mathrm{x}+1}\ \circ\ \EuScript{R}^{-1}\ \circ\ \EuScript{T}^{-1},   \label{eq:coef:ass}\end{equation}
and scheduling region  $\mathbb{P}=\EuScript{T}(\Y_0 \times \U_0)$ satisfies 
\begin{equation} \label{eq:behav:SS1}
\mathfrak{C}_\mathrm{SS}^\mathrm{o}  \subseteq \pi_p \mathfrak{B}_\mathrm{SS}^\mathrm{o},  
\end{equation}\end{subequations}
where
 \begin{multline*}
\pi_p \mathfrak{B}_\mathrm{SS}^\mathrm{o}=\bigl\{ (u,x,y)\in  \mathcal{C}_{n_\mathrm{x}}(\R,\U_0\times\mathbb{X}_0 \times \Y_0) \mid 
\exists p \in  \mathcal{C}_{n_\mathrm{x}}(\R,\mathbb{P}), 
\exists z \in  \mathcal{C}_{n_\mathrm{x}}(\R,\Y_0^{n_{\mathrm x}}) \text{ such that (\ref{eq:ch3:03}--b) hold} \\
 \text{while $x=\Psi(z,u,  \ldots, \tfrac{d^{n_{\mathrm x}}}{dt^{n_{\mathrm x}}} u)$}  \bigr\}, 
\end{multline*}
and
\begin{equation*}
\mathfrak{C}_\mathrm{SS}^\mathrm{o}=\bigl\{ (u,x,y) \in  \mathcal{C}_{n_\mathrm{x}}(\R,\U_0\times\mathbb{X}_0 \times\Y_0) \text{ such that (\ref{e:nl1}--b) hold} \bigr\}.
\end{equation*}
\end{theorem}
In terms of Theorem \ref{thm:1}, the set of all $(\U_0 \times \X_0 \times \Y_0)$ admissible solutions  of \eqref{e:nl} can  be embedded into the solution set of an LPV-SS representation and \eqref{eq:sche:choice1} gives a direct selection of the scheduling variables under the factorization w.r.t. $\{\phi_i\}_{i=1}^{\tau}$.
\begin{proof}
  Consider a
    $(\U_0\times\mathbb{X}_0 \times\Y_0)$  admissible solution  $(u,x,y)$  of \eqref{e:nl} and  invoke the definitions \eqref{eq:zv}. Let $\xi=[\begin{array}{cc} z^\top & v^\top \end{array}]^\top$ and $\zeta=\left[\tiny \begin{bmatrix} y \\ u \end{bmatrix} , \frac{d}{dt} \begin{bmatrix} y \\ u \end{bmatrix}, \ldots, \frac{d^{n_\mathrm{x}-1}}{dt^{n_\mathrm{x}-1}} \begin{bmatrix} y \\ u \end{bmatrix}\right] $. Notice that $\zeta=\EuScript{R}\xi$ and $\xi=\EuScript{R}^{-1}\zeta$. Introduce $\EuScript{P}$ and $\EuScript{P}^{-1}$ which are $n_\mathrm{x}$-times block diagonal matrices of $\EuScript{T}$ and $\EuScript{T}^{-1}$, respectively.
     Notice that  $\EuScript{P} \EuScript{P}^{-1}\EuScript{P}\EuScript{R}\xi = \EuScript{P}\EuScript{R}\xi$ and hence
$\EuScript{P} (\EuScript{R} \xi - \EuScript{P}^{-1}  \EuScript{P}\EuScript{R}\xi)=0$.
 From the definition of the selection matrices
it follows that
\[
   r_i\ \circ\ \EuScript{R}^{-1}(\EuScript{R} \xi) =
   r_{i}\ \circ\ \EuScript{R}^{-1}(\EuScript{P}^{-1}  \EuScript{P} \EuScript{R}\xi)= 
    r_{i}\ \circ\ \EuScript{R}^{-1}\ \circ\  \EuScript{T}^{-1} (\EuScript{P}  \EuScript{R} \xi) .\]
Define $\tilde{p}=[\begin{array}{cc} y^\top & u^\top \end{array}]^\top$. 
Notice that 
\[ 
  \EuScript{P} \EuScript{R} \xi = \EuScript{P}  \zeta = \begin{bmatrix} (\EuScript{T} \tilde{p})^\top & \ldots & \frac{d^{n_{\mathrm x}-1}}{dt^{n_{\mathrm x}-1}} (\EuScript{T} \tilde{p})^\top \end{bmatrix}^\top=\begin{bmatrix} p^\top & \ldots & \frac{d^{n_{\mathrm x}-1}}{dt^{n_{\mathrm x}-1}} p^\top \end{bmatrix}^\top.
\]
Hence, 
\begin{equation*}
    r_i(\xi)= r_i\ \circ\ \EuScript{R}^{-1}\ \circ\ \EuScript{T}^{-1}(\EuScript{P}\EuScript{R} \xi) =
    \left\{\begin{array}{rl} \alpha_{i-1} \diamond  p,  & i\in\mathbb{I}_1^{n_{\mathrm x}}; \\ \beta_0 \diamond p, &  i=n_{\mathrm x}+1. \end{array}\right.
\end{equation*}
From the discussion above and using $\frac{d}{dt} z_{i} = z_{i+1}$ for $i=\mathbb{I}_1^{n_{\mathrm x}-1}$ it follows that 
 \begin{equation}
    \frac{d^{n_\mathrm{x}}}{dt^{n_\mathrm{x}}}z_{n_\mathrm{x}}= \Gamma_{n_{\mathrm x}}\left(\begin{bmatrix} y \\ u \end{bmatrix}, \ldots, \frac{d^{n_\mathrm{x}-1}}{dt^{n_\mathrm{x}-1}} \begin{bmatrix} y \\ u \end{bmatrix}\right)  
      = \sum_{i=1}^{n_{\mathrm x}} r_i(\xi)z_{i} + r_{n_{\mathrm x}+1}(\xi)u  
      = \sum_{i=0}^{n_{\mathrm x}-1} (\alpha_i \diamond p) z_{i+1} +(\beta_0 \diamond p)u.  \label{eq:pr1}
   \end{equation}
 Hence, $(u,z,y,p)$ is a solution of the LPV-SS representation  \eqref{e:lpv_obvs_n} defined in the statement of the theorem.
 Moreover, since $\Psi$ is a $(\U_0\times\mathbb{X}_0 \times\Y_0)$ observability function and \eqref{eq:pr1} holds, 
 $x=\Psi(z,u, \ldots, \frac{d^{n_\mathrm{x}-1}}{dt^{n_\mathrm{x}-1}} u)$.  
\end{proof}

In order to make Theorem \ref{thm:1} applicable, we need an algorithm to compute the factorization of the function $\Gamma_{n_\mathrm{x}}\circ\ \EuScript{R}$ on $\mathbb{V}=\Y_0^{n_{\mathrm x}} \times \U_0^{n_{\mathrm x}}$
 with respect to $\{\phi_i: \mathbb{V} \rightarrow \R\}_{i=1}^{\tau}$. Let $N$ and $D$ be such polynomials that $\Gamma_{n_\mathrm{x}}\circ\ \EuScript{R}$  can be written as \eqref{eq:ratform}.
Then, Algorithm \ref{alg:fact1}, which takes $N$ and $D$ and $\{\phi_i\}_{i=1}^{\tau}$ as parameters, returns a factorization 
$(\{r_{i}\}_{i=1}^{m},s)$ of $\Gamma_{n_\mathrm{x}}\circ\ \EuScript{R}$ with respect to the first $m=n_\mathrm{x}+1$ variables, \emph{i.e.}, $\{z_{i}=\frac{d^i}{dt^i}y\}_{i=1}^{n_\mathrm{x}}$ and $u$.

\begin{algorithm}
\caption{Factorization}\label{alg:fact1}
\begin{algorithmic}
\Require {$N(X_1,\ldots, X_{n+\tau}) ,D(X_1,\ldots,X_{n+\tau}), \{\phi_i\}_{i=1}^{\tau}$, $m \le n$}
\State  $S \gets N$. 
\For{$k \gets m:1$}
\State  represent $S$ as $ \sum_{(i_1,\ldots, i_{n+\tau}) \in \mathbb{I}} \gamma_{i_1,\ldots,i_{n+\tau}} X_1^{i_1} \cdots X_{n+\tau}^{i_{n+\tau}}$ for a finite  index set $\mathbb{I} \subseteq \mathbb{N}^{n+\tau}$. \vspace{1mm}
\State $M_k \gets \sum_{(i_1,\ldots, i_{n+\tau}) \in \mathbb{I}, i_k \ge 1} \gamma_{i_1,\ldots,i_{n+\tau}} \frac{X_1^{i_1} \cdots X_{n+\tau}^{i_{n+\tau}}}{X_k}$. \vspace{1mm}
\State $S\gets S-M_k X_k$. 
\EndFor \vspace{1mm}
\State $r_i(\xi) \gets \frac{M_i(\xi,\phi_1(\xi),\ldots, \phi_{\tau}(\xi))}{D(\xi,\phi_1(\xi),\ldots,\phi_{\tau}(\xi))}$,\ \ $s(\xi) \gets \frac{S(\xi,\phi_1(\xi),\ldots, \phi_{\tau}(\xi))}{D(\xi,\phi_1(\xi),\ldots,\phi_{\tau}(\xi))}$,\ \ $\xi \in \mathbb{V}$. \vspace{1mm}
\State
\Return $(\{r_i\}_{i=1}^{m},s)$. 
\end{algorithmic} 
\end{algorithm}

Theorem \ref{thm:1} indicates that it is possible to embed NL systems into LPV-SS representations in a systematic way.  
Furthermore, it characterizes an LPV embedding in terms of a multi-path linearization which resembles  feedback linearization of NL systems. However,  in feedback linearization,  a virtual input signal is introduced so that the transformed system becomes LTI. In contrast, in the  proposed LPV approach, a set of virtual variables, denoted by $p$, are constructed  which result in a varying linear relationship.  
Thus, the obtained LPV-SS representation is  useful to develop controllers that can shape the closed-loop behavior unrestricted or have better robustness than with an LTI target behavior.
Furthermore, $p$ is selected to be state-independent (in contrast with the common NL to LPV conversion techniques) meaning that in practice, the LPV controller designed for this model can be directly applied in a real-world system. Furthermore, the dimension of $p$ is reduced by considering the essential support of $\{r_i\}_{i=1}^{n_\mathrm{x}+1}$. 
On the other hand, Theorem \ref{thm:1}  guarantees the embedding and hence the validity of the LPV representation only
for those state trajectories $x$ of the NL system which remain in $\X_0$ and for those inputs $u$ which remain in $\U_0$.
Hence, when designing controllers using the LPV-SS form, one must ensure that $u(t)\in \U_0$ and $x$ remains in $\X_0$. For the latter, it is enough to ensure that the state $z$ of the 
LPV-SS model remains in $\Y_0^{n_{\mathrm x}}$. Otherwise, the LPV-SS representation of the NL system is no longer valid.  

\subsection{Choice of the scheduling variable}\label{sec:schechoice}
Although Theorem \ref{thm:1} gives a straightforward formulation of the LPV-SS representation of \eqref{e:nl} with a unique choice of $p$, one may consider projections of this variable to simplify the resulting dependency structure of \eqref{e:nl} as follows:

%%%%%%%%%%%%%%%%%%%%%%
\begin{itemize}
\item  \emph{Full dynamic dependency}: \eqref{eq:sche:choice1} results in a possible dynamic dependence of \eqref{eq:dep} on $p= \EuScript{T} \mas{cc}{y & u}^\top$ with $\mathbb{P}=\EuScript{T}(\Y_0 \times \U_0)\subseteq \mathbb{R}^m$, $m\leq n_\mathrm{y}+n_\mathrm{u}$,  characterized by rational combinations of the chosen $\{\phi_i\}_{i=1}^{\tau}$. Although such a choice is tempting from the theoretical and even identification point of view, as it minimizes the conservativeness of the embedding, it results in models which are difficult for control design. Current techniques are only able to handle rational static dependence on $p$.
\item \emph{Rational dependence}: Using the ``minimal'' scheduling choice characterized by Theorem \ref{thm:1}, it is possible to introduce a so-called \emph{scheduling map} $\eta$: 
\begin{equation}
p=\eta\diamond (y,u)=\left[\begin{array}{cccc} \EuScript{T} \mas{cc}{y & u}^\top & \phi_1( \EuScript{T} \mas{cc}{y & u}^\top ,\ldots,\frac{d^{n_\mathrm{x}-1}}{dt^{n_\mathrm{x}-1}} \EuScript{T} \mas{cc}{y & u}^\top )  &\ldots & \phi_\tau( \EuScript{T} \mas{cc}{y & u}^\top ,\ldots,\frac{d^{n_\mathrm{x}-1}}{dt^{n_\mathrm{x}-1}} \EuScript{T} \mas{cc}{y & u}^\top ) \end{array} \right]^\top.
\end{equation}
Hence, by increasing $\mathrm{dim}(p)$ to $m+\tau$, where $m$ is the number of rows in $\EuScript{T}$, the dynamic nature of the dependence can be hidden 
into $\eta$ and the $p$-dependence of \eqref{eq:dep} is reduced to be static rational. This is desirable for control and identification as $\eta$ can be applied on the measured values of $(u,y)$ to compute $p$. Note that increasing the dimensions of $p$ leads to more conservatism as $\pi_p \mathfrak{B}_\mathrm{SS}^\mathrm{o} $ grows with every hidden relation in $\eta$.
\item \emph{Affine dependence}: The previous procedure can also be applied to hide even the polynomial dependence resulting from the above mentioned procedure by constructing a map $p=\eta\diamond (y,u)$ which, by substituting it to \eqref{eq:coef:ass}, results in an affine dependence of \eqref{eq:dep} on $p$. While this is tempting to simplify control synthesis  based on such an embedding, it also maximizes the conservativeness of $\pi_p \mathfrak{B}_\mathrm{SS}^\mathrm{o} $.
\end{itemize}

\noindent Note that  computation of the analytic map $\Gamma_{n_\mathrm{x}}$ requires inversion of functions, and hence in general, it is not guaranteed that it has a closed form. 
While theoretically this does not hinder the application of Theorem \ref{thm:1}, it makes the calculation of the LPV model described in Theorem \ref{thm:1} far from trivial.  In principle, what is required for Theorem \ref{thm:1} is not an analytic expression for $\Gamma_{n_\mathrm{x}}$, but an expression for the factorization of $\Gamma_{n_\mathrm{x}}$. The latter might be computable even if there is no analytic expression for $\Gamma_{n_\mathrm{x}}$.

In conclusion, Theorem  \ref{thm:1} reveals that LPV embedding of an NL system is affected by a trade-off between conservativeness and the simplicity of dependence of the resulting representation on $p$. 
In this respect, it is interesting to observe that the choice of basis functions $\{\phi_i\}_{i=1}^\tau$ does not influence the validity of the transformation nor the controllability or observability of the resulting model as long as there is no remainder term, \emph{i.e.}, $s=0$. %The resulting $r_i$ functions are invariant w.r.t. the choice of $\{\phi_i\}_{i=1}^\tau$. 
However, when $\{\phi_i\}_{i=1}^\tau$ are absorbed into $\eta$, their choice has a significant impact on the conservativeness of the embedding. As in system identification, the choice of $\eta$ is invisible for the estimation procedure and it can seriously affect the outcome of the estimation (persistency of excitation, correlation with noise, etc.), while in control, robustness of the control law can be analyzed against variations of the LPV-SS representation, but not against variations in $\eta$. Additionally, in LPV-MPC, hidden relations in $\eta$, especially dependence on $u$, can seriously compromise the meaningfulness of the resulting optimization problem; hence, in principle, control design and LPV model development, in terms of the choice of $\eta$ should be seen as a joint process, see \cite{Hoffmann2015,Hoffmann2015b}.

\subsection{Handling the remainder term}\label{sec:trimming}

Theorem \ref{thm:1} deals with the case when $s=0$, \emph{i.e.}, $\Gamma_{n_{\mathrm x}}$ can be factorized without a remainder. Suppose that the
conditions of Theorem \ref{thm:1} hold, but $s\neq 0$. In this case, we can still represent the solutions of \eqref{e:nl} by
solutions of an LPV system (similarly to Theorem \ref{thm:1}), but the resulting representation will not be linear due to the extra $p$-dependent affine term $\gamma := s\ \circ\ \EuScript{R}^{-1}\ \circ\ \EuScript{T}^{-1}$. 
This term is
undesirable both in LPV control synthesis and identification as the  whole LPV framework builds upon the assumed linearity of the system description. As this phenomenon is not uncommon in applied LPV control, we collected here the possible strategies to deal with affine terms: 
\begin{itemize}
\item \emph{Virtual input}: An input-disturbance signal $ d \equiv1$ is introduced to incorporate the affine term into the $\EuScript{B}$ matrix: 
$$  \tilde{\EuScript{B}}\diamond p= \bbma
    0 & 0\\
    \vdots & \vdots\\
    0 & 0\\
     (\beta_{0}\diamond p) & (\gamma\diamond p)
    \ebma \quad \text{with new input: } \bbma  u \\ d \ebma . $$
 Then, considering $d$ as a time-varying disturbance with an $\mathcal{L}_2$ norm bound of $1$, optimal control synthesis or MPC control can be conveniently applied. Although this strategy changes the IO partition of the system and it increases the conservativeness of the embedding, it leads to a complete representation of the original NL  behavior.
\item \emph{Ignored} in the LPV ``representation'' of the system behavior and during  control synthesis one of the following choices are applied
  \begin{itemize}
  \item The designed controller is augmented with a feedforward path to compensate for $\gamma$ during control implementation, see \cite{HaGuWe13, RuKrHoBe08}.
    \item Input disturbance rejection is considered as a control objective.
    \end{itemize} 
      \item \emph{Enforced factorization}: $\gamma$ is rewritten as $\frac{\tilde{\gamma}}{u}u$ or $\frac{\tilde{\gamma}}{z_j}z_j$ and added to $\beta_0$ or $\alpha_j$, respectively. The associated $u$ or $z_j$ should never approach close to the origin during operation, otherwise loss of stability might occur, see \cite{Kwiatkowski06,Toth2010SpringerBook} for more details.\end{itemize}

 \subsection{Scheduling with signal derivatives}\label{sec:trimming}
 
Using the the proposed model conversion method, $p$ can potentially contain $n_\mathrm{x}-1$ time derivatives of $(y,u)$. To implement an LPV controller $\mathcal{K}$ designed with the resulting model, derivatives of $u$ correspond to derivatives of the output of $\mathcal{K}$, which can be obtained by an extended state realization of $\mathcal{K}$. Regarding derivatives of $y$, the following options are available:
\begin{itemize}
\item \emph{Direct measurement:} In many \MOD{mechatronic} applications, the underlying IO relations are $2^\mathrm{nd}$-order in nature and often velocity and acceleration measurements are available (just think about IMUs in ground and aerial vehicles or flowmeters, rotameters and a huge array of various designs of gyroscopes and accelerometers).
\item \emph{Numerical differentiation} and filtering methods, designed to mitigate the effect of noise and approximation error on the derivatives (see e.g., \cite{Atkinson78, Shao2003157,Savitzky1964,Pintelon65799,Rabiner1162090,Ferrer2013}), can be used to calculate the derivatives of $y$. 
\item \emph{Observer design:} The model of the plant dynamics can be transformed to an observability form where the state variables directly correspond to the derivatives of $y$ up to the relative degree of the system and the rest of the state variables can be used to compute higher derivatives of $y$ when the derivatives of $u$ are known. This means that derivatives of $y$ can be estimated by an observer or a Kalman filter as any other state variables. Commonly derivatives of $y$ naturally appear among the state variables of first-principles based plant models, like position, velocity, acceleration in motion equations of mechanical systems.
\end{itemize}
When identification of the resulting LPV model is considered, in continuous time, computation of time-derivatives of $(y,u)$ in either frequency domain or the time-domain, in prediction or simulation are required by most identification methods (subspace methods, prediction-error minimization, instrumental variables, etc.). Therefore, handling derivatives of $(y,u)$ is a natural step in many cases, only the means of obtaining them differs.
 
Compared to the proposed conversion method, alternative conversion methods to LPV form often choose state-variables of the NL model in an ad-hoc manner to be part of $p$. With such a choice, $p$ is often not measurable and the LPV controller $\mathcal{K}$ has to be used together with an observer for estimating $p$. However, $\mathcal{K}$ was designed with the assumption that $p$ is known. Hence, by introducing an observer for estimating $p$,  the stability and performance guarantees of the LPV controller are lost. Of course, one can argue that delay and performance loss can also be introduced with numerical differentiation or filtering methods in case derivatives of $y$ are not directly measurable. In that case, we run into the same problem with the proposed methodology. In fact, the same choice occurs in feedback linearization when one can choose between using derivatives of $y$ or the states $x$ or the original system to calculate the linearizing feedback. The proposed methodology in this paper aims at providing systematic options beyond using only $x$ in the scheduling map.
 
\subsection{Computation of $\Gamma_{n_{\mathrm x}}$}
\label{sect:rel:degree} 
For the sake of completeness, the construction procedure of $\Gamma_{n_\mathrm{x}}$, which is used in Theorem \ref{thm:1}  and relies on known NL system theory concepts, is presented next. \begin{definition}[Relative degree \cite{Is95}]
\label{nl_reldegree}
The NL-SS system representation \eqref{e:nl} is said to have relative degree $n_\mathrm{r}$ at a point $x_0\in\mathbb{X}$ if there exists an open subset $x_0\in\mathbb{X}_\mathrm{r}  \subseteq \mathbb{X}$ such that
\begin{itemize}
  \item[(i)] $L_g L_f^i h(x)=0$, \ $\forall x \in \mathbb{X}_\mathrm{r}$, \ $i<n_\mathrm{r}-1$,
  \item[(ii)] $L_g L_f^{n_\mathrm{r}-1}h(x_0)\ne 0$,
\end{itemize}
where $L_f^{i}h(\cdot)$ stands for the $i^\m{th}$ Lie-derivative of $h$ w.r.t.\ $f$.
\end{definition}
Note that not every NL system represented in the form of \eqref{e:nl} has a relative degree $n_\mathrm{r}$ at all. Neither is it true that the same $n_\mathrm{r}$ qualifies  for all $x_0\in\mathbb{X}$.  We refer to \cite{Is95} for more in depth discussion on this topic. In the sequel,  it is assumed that $x_0$ is chosen such that the relative degree of  \eqref{e:nl} is well-defined at this point. Next, we consider the construction of $\Gamma_{n_{\mathrm x}}$ in a neighborhood of $x_0$ in two cases: when $n_\mathrm{r}$ of \eqref{e:nl}  at $x_0$ equals $n_\mathrm{x}$ and 
when $n_\mathrm{r}<n_{\mathrm  x}$. 

%-------------------------------------------------------------------------------
\subsubsection{Case of $n_\mathrm{r}=n_\mathrm{x}$}\label{ss:r=n}
%-------------------------------------------------------------------------------

Consider a solution $(u,x,y)$ of \eqref{e:nl}, such that for all $t \in \mathbb{R}$, $x(t) \in \mathbb{X}_\mathrm{r}$ (see Definition \ref{nl_reldegree}).  
In this case,  
\begin{subequations}\label{e:nl_new_r=n}
\begin{align}
        z_1&= y = h(x)=\Phi_0(x), \\[-1mm]
                & \hspace{0.5mm}  \vdots   \notag \\[-2mm]
        z_{n_\mathrm{x}}&=  \frac{d^{n_\mathrm{x}-1}}{dt^{n_\mathrm{x}-1}}y = L_f^{{n_\mathrm{x}}-1} h(x)=\Phi_{{n_\mathrm{x}}-1}(x),
          \end{align}
          while
          \begin{equation}
           \frac{d^{n_\mathrm{x}}}{dt^{n_\mathrm{x}}}y =L_f^{n_\mathrm{x}} h(x) + L_g L_f^{{n_\mathrm{x}}-1} h(x) u = \Phi_{n_\mathrm{x}}(x,u),
          \end{equation}
 \end{subequations}
\emph{i.e.}, only the $n_{\mathrm x}^\mathrm{th}$ derivative of $y$ depends on $u$. This gives  
 \begin{equation} \label{eq:ss:map}
 z= \Phi(x)=\bbma
    h(x) & L_f h(x) & \dots  & L_f^{n_\mathrm{x}-1} h(x)
   \ebma^\top,
\end{equation}
hence the local inverse of $\Phi$ provides the observability function $\Psi$ in Definition \ref{e:nl:obs} to construct $\Gamma_{n_{\mathrm x}}$. Recall \cite[Lemma 4.1.1, p. 140]{Is95}\, that 
 if the relative degree $n_\mathrm{r}$ of \eqref{e:nl} is $n_\mathrm{x}$ at $x_0$, then the gradients 
\(\nabla h(x_0), \dots, \nabla L_f^{n_\mathrm{x}-1} h(x_0) \) are linearly independent. Hence, in this case,
 the Jacobian of $\Phi(x_0)$ is invertible based on the  
 \emph{inverse function theorem} \cite{DieudonneBook}:
\begin{lemma}[Inversion of $\Phi$]
\label{p:diffeomorphism_r=n}
There exist open sets $\X_0 \subseteq \mathbb{X}_\mathrm{r}$ and  $\Y_0 \subseteq \R$,  such that $x_0 \in \X_0$, $\Phi(\X_0)=\Y_0^{n_{\mathrm x}}$  and $\Phi|_{\X_0}: \mathbb{X}_0 \rightarrow \Y^{n_{\mathrm x}}_0$,
the restriction of  $\Phi$ to $\X_0$, is an analytic diffeomorphism, \emph{i.e.}, the analytic inverse $\Phi^{\dag}$ of $\Phi|_{\X_0}$ exists. 
\end{lemma}

By a slight abuse of notation, we will identify $\Phi$ in the sequel with its restriction to $\X_0$, \emph{i.e.}, we will view it as a diffeomorphism 
$\Phi|_{\X_0}: \mathbb{X}_0 \rightarrow \Y^{n_{\mathrm x}}_0$.
Let $\U_0$ be an arbitrary open subset of $\U$.
We can define the observability map $\Psi:(\Y_0 \times \U_0)^{n_{\mathrm x}} \rightarrow \X_0$, satisfying Definition \ref{e:nl:obs}, by
\begin{equation} \label{eq:cont1:psi} \Psi\!\left( \! \begin{bmatrix} \mu_1 \\ \upsilon_1 \end{bmatrix}, \ldots \begin{bmatrix} \mu_{n_{\mathrm x}} \\ \upsilon_{n_{\mathrm x}} \end{bmatrix} \!\right)
=\Phi^{\dag}\!\left(\! \mu_1,\ldots,\mu_{n_{\mathrm x}} \! \right),\end{equation}
for all $\mu_1,\ldots,\mu_{n_{\mathrm x}} \in \Y_0$ and $\upsilon_1,\ldots,\upsilon_{n_{\mathrm x}} \in \U_0$.
Note that, in this case, $\Psi$ does not depend on $\{\upsilon_i\}_{i=1}^\mathrm{n_\mathrm{x}}$. 
Hence, by the construction in Theorem \ref{thm:1}, $ \Gamma_{n_{\mathrm x}}$ results in
\begin{equation} \label{eq:state_trafo2}
 \Gamma_{n_{\mathrm x}}\left(\begin{bmatrix} \mu_1 \\ \upsilon_1 \end{bmatrix}, \ldots, \begin{bmatrix} \mu_{n_{\mathrm x}} \\ \upsilon_{n_{\mathrm x}} \end{bmatrix}\right) = 
L_f^{n_\mathrm{x}}h(\Phi^{\dag}(\mu_1,\ldots,\mu_{n_{\mathrm x}}))+  L_gL^{n_\mathrm{x}-1}_fh(\Phi^{\dag}(\mu_1,\ldots,\mu_{n_{\mathrm x}}))\upsilon_1 . \end{equation} 
%-------------------------------------------------------------------------------
\subsubsection{Case of $n_\mathrm{r}<n_\mathrm{x}$}\label{ss:r<n_1}
%-------------------------------------------------------------------------------
Computing $z_1=y,\dots,z_{n_\mathrm{r}}=\frac{d^{n_\mathrm{r}-1}}{dt^{n_\mathrm{r}-1}}y$ follows as in \eqref{e:nl_new_r=n}, but 
\begin{equation}\label{e:nl_new_r<n_2}
z_{n_\mathrm{r}+1}  =  \frac{d^{n_\mathrm{r}}}{dt^{n_\mathrm{r}}}y  = 
L_f^{n_\mathrm{r}} h(x) + L_g L_f^{n_\mathrm{r}-1} h(x) u = \Phi_{n_\mathrm{r}}(x,u).
\end{equation}
Continuing the construction of the map gives that
\begin{equation}\label{e:d+1}
z_{n_\mathrm{r}+2}=\frac{d}{dt}z_{n_\mathrm{r}+1} =  
            L_f^{n_\mathrm{r}+1} h(x) + L_g L_f^{n_\mathrm{r}} h(x) u  + L_f L_g L_f^{n_\mathrm{r}-1} h(x) u   + L_g^2 L_f^{n_\mathrm{r}-1} h(x) u^2 + L_g L_f^{n_\mathrm{r}-1} h(x) \frac{d}{dt}u=  
      \Phi_{n_\mathrm{r}+1}(x,u, \tfrac{d}{dt} u). 
\end{equation}
Repeating this operation recursively results in
\begin{equation} \label{eq:ext:ssmap}
\frac{d}{dt}z_{n_\mathrm{r}+l} =\Phi_{n_{\mathrm r} + l-1} (x,u,\ldots, \tfrac{d^{ l-1}}{dt^{l-1}}u), 
\end{equation}
for $1\leq l\leq  n_\mathrm{s} +1$ with $ n_\mathrm{s}= n_\mathrm{x}-n_\mathrm{r}-1$. Compared to the previous case, these maps now depend on $u,\ldots \frac{d^{n_\mathrm{s}}}{dt^{n_\mathrm{s}}}u$. Hence,  
 \begin{equation} \label{eq:ss:map2}
 z= \Phi(x,u,\ldots, \tfrac{d^{n_{\mathrm s}}}{dt^{n_{\mathrm s}}}u)=
\left[\begin{array}{cccccc} \Phi_0(x) &\ldots & \Phi_{n_{\mathrm r}-1}(x)& \Phi_{n_{\mathrm r}}(x,u)&  \ldots &
\Phi_{n_{\mathrm x}-1}(x,u,\ldots, \frac{d^{n_{\mathrm s}}}{dt^{n_{\mathrm s}}}u) \end{array}\right]^\top,
\end{equation}
and the local inverse of $\Phi$ provides $\Psi$ in Definition \ref{e:nl:obs} to construct $\Gamma_{n_{\mathrm x}}$.
 We can now state the following lemma presenting the conditions for local invertibility of $\Phi$. 
\begin{lemma}[$\Phi$ inversion under $n_\mathrm{r}<n_\mathrm{x}$]
\label{ss:rn_1:lemma1} Assume full rank of $\nabla \Phi (x_0, u_0, \ldots u_0)$, where $\nabla \Phi$ is the Jacobian of $\Phi$ w.r.t.~$x$. %, i.e. $\nabla \Phi(x,u_1,..,u_l)=(\frac{dPhi_i}{dx_j}(x,u_1,..u_l))_{i,j=1}^{n_x}$. 
There exist open sets $x_0\in \mathbb{X}_0 \subseteq \mathbb{X}_\mathrm{r}$, $u_0\in \mathbb{U}_0 \subseteq \R$,
  $\mathbb{Y}_0 \subseteq \R$, and an analytic function
  $\Phi^{\dag}: \Y_0^{n_{\mathrm x}} \times \U_0^{n_{\mathrm s}+1} \rightarrow \mathbb{X}_0$, such that for all
  $\mu \in  \Y_{0}^{n_{\mathrm x}}, \upsilon  \in  \U_0^{n_{\mathrm s}+1}$ and  \vspace{-0.5mm}
  $x \in \X_0$: 
  \[ \mu =\Phi (x, \upsilon) \iff  x=\Phi^{\dag}(\mu,\upsilon). \] 
\end{lemma}
Lemma \ref{ss:rn_1:lemma1} follows from the implicit function theorem\cite{DieudonneBook} applied to $\mu-\Phi(x,\upsilon)$. Using $\Phi^{\dag}$, we can 
 define the function $\Psi: (\Y_0 \times \U_0)^{n_{\mathrm x}} \rightarrow \X_0$ similarly as in \eqref{eq:cont1:psi} which satisfies Definition \ref{e:nl:obs} by construction.  Then, we can proceed with the construction of  $\Gamma_{n_{\mathrm x}}$ as in Definition \ref{e:nl:obs:cor},
 except that  $\Gamma_{n_{\mathrm x}}$ will not depend on the last $n_{\mathrm r}$ components of $\U^{n_{\mathrm x}}$ and hence it can be defined on
 $\Y_0^{n_{\mathrm x}} \times \U_0^{n_{\mathrm s}+1}$ instead of $\Y_0^{n_{\mathrm x}} \times \U_0^{n_{\mathrm x}}$.

The price to pay for a system with relative degree less than its order is that the resulting LPV model through $p$ depends on $u$ and its derivatives up to order  $n_\mathrm{s}$. On the other hand, all scheduling signals are directly computable form measured variables without requiring  the original states of the system. 

%<<<<<<<<<<<<<<<<<<<<<<<<<<<<<<<<<<<<>>>>>>>>>>>>>>>>>>>>>>>>>>>>>>>>>>>>>>>>>>

\section{Conversion  to the full observability form}\label{s:nl2lpvobvs}
%<<<<<<<<<<<<<<<<<<<<<<<<<<<<<<<<<<<<>>>>>>>>>>>>>>>>>>>>>>>>>>>>>>>>>>>>>>>>>>

One of the shortcomings of the conversion procedure of  Section \ref{s:nl2lpvobvs_n} 
is that in case of relative degree $n_\mathrm{r}< n_\mathrm{x}$, the conversion results in an LPV model ``depending'' on $\{\frac{d^l}{dt^l}u \}_{l=0}^{n_\mathrm{s}}$. This dynamic dependence on $u$ can be undesirable as it increases the complexity of the resulting model.  One can say that this is the price to be paid for trying to use only $\beta_0$ to express the relations involving $u$. 
One way to overcome this is to assume that a part of the state is available for measurement. In that case,
parts of $x$ become components of $p$ and they are used to replace the derivatives of $u$ in the dependency structure.

To gain some intuition,  consider  \eqref{e:nl} with a well-defined relative degree $n_\mathrm{r}<n_\mathrm{x}$ at a point $x=x_0$ and the transformation map $\Phi$ from \eqref{eq:ss:map}.  If $(u,x,y)$ is a solution of \eqref{e:nl} such that for all $t\in\mathbb{R}$, $x(t) \in \mathbb{X}_\mathrm{r}$, then, even in case of $n_\mathrm{r}<n_\mathrm{x}$, it is possible to use $z(t) = \Phi(x(t))$ as the state of the LPV model, constructed as \vspace{-0.5mm}
\begin{equation} \label{eq:dep:full12}
 z=\mas{cccccc}{y & \cdots & \frac{d^{n_\mathrm{r}-1}}{dt^{n_\mathrm{r}-1}}y & L_f^{n_\mathrm{r}}h(x) & \cdots & L_f^{n_\mathrm{x}-1}h(x)  }.
\end{equation}
 Notice that
\begin{equation} \label{eq:bbb}
\frac{d}{dt}z_{n_\mathrm{r}+l} =\underbrace{ L_f^{n_\mathrm{r}+l} h(x)}_{z_{n_\mathrm{r}+l-1}} + L_g L_f^{n_\mathrm{r}+l-1} h(x) u, \vspace{-0.5mm}
\end{equation}
for $1\leq l\leq n_\mathrm{s}+1$. In terms of Lemma \ref{p:diffeomorphism_r=n}, there exist open sets $x_0\in\X_0 \subseteq \mathbb{X}_\mathrm{r}$, $\Y_0 \subseteq \mathbb{R}$,  such that $\Phi|_{\X_0}:\X_0 \rightarrow \Y_0^{n_{\mathrm x}}$ is an analytic diffeomorphism with analytic inverse $\Phi^\dag$. %Let $\U_0$ be any open subset of $\R$ containing $0$. 
Hence, for $x(t)\in\mathbb{X}_0$, $\forall t \in\mathbb{R}$ and each $l$, we can write the $u$-related terms in \eqref{eq:bbb} as 
\begin{equation}
L_g L_f^{n_\mathrm{r}+l-1} h(x) u =\underbrace{ L_g L_f^{n_\mathrm{r}+l-1} h(\Phi^\dag(z) )}_{\beta_{n_\mathrm{s}+1-l}} u.
\end{equation}
This implies that the last state equation reads as
\begin{equation}
   \frac{d}{dt}z_{n_\mathrm{x}} = \underbrace{L_f^{n_\mathrm{x}}h(\Phi^{\dag}(z) )}_{\Gamma_{n_{\mathrm x}}(z)}+\underbrace{L_gL^{n_\mathrm{x}-1}_fh(\Phi^{\dag}(z))}_{\beta_0(z)}u.  \label{eq:last:eq1}
\end{equation}

Intuitively, we want to factorize $\Gamma_{n_\mathrm{x}}  (z)$, \emph{i.e.}, write it as
$\Gamma_{n_\mathrm{x}}  (z)=\sum_{i=0}^{n_{\mathrm x}-1} \alpha_i(z)z_{i+1}$. As a result, we obtain Equations (\ref{eq:ch3:03}--b) where $\{\alpha_i\}_{i=0}^{n_\mathrm{x}-1}$ and $\{\beta_i\}_{i=0}^{n_\mathrm{s}}$ are dependent on $z$. Using that $z$ satisfies \eqref{eq:dep:full12}, which is dependent on $\{\frac{d^l}{dt^l} y\}_{l=0}^{n_\mathrm{r}-1}$ and $x$, we arrive at an LPV model by taking\footnote{In theory, it is possible to consider $p=x$. However, this choice results in a scheduling region as large as $\mathbb{X}$ and the resulting LPV model will be overly conservative. Hence, it is a better strategy to include $y$ into $p$ since the derivatives of $y$ and  $x$ are closely related. We would then hope that in the final LPV model, most of the state components disappear from $p$.}
$p$ as a linear projection of 
$[\ y^\top \ \ x^\top\ ]^\top$.

Now, we present the  above procedure more formally. 
Define the maps $\Gamma_{n_\mathrm{x}}: \Y_0^{n_{\mathrm x}}  \rightarrow \mathbb{R}$ and  $\{ \tilde{\beta_i}: \Y_0^{n_{\mathrm x}} \rightarrow \mathbb{R}\}_{i=0}^{n_\mathrm{x}-1}$ as
\begin{subequations}
\begin{align} 
\label{eq:last:eq}
  \Gamma_{n_\mathrm{x}}(\zeta)&=L_f^{n_\mathrm{x}}h(\Phi^{\dag}(\zeta)), \\
  \tilde{\beta}_i(\zeta)& = \left\{\begin{array}{cl} 
                                        L_gL_f^{n_{\mathrm x} - i-1}h(\Phi^{\dag}(\zeta)) &\ \ i \le n_{\mathrm s}; \\
                                        0 &\ \ \mbox{otherwise;}
                                      \end{array}\right.\label{eq:last:eq_betas}
\end{align}
\end{subequations}
for all $\zeta \in \Y_0^{n_{\mathrm x}}$. 
Assume that there exists a set of analytic functions $\{\phi_i\}_{i=1}^{\tau}$ on $\mathbb{Y}^{n_{\mathrm x}}_0$
such that the map
$\Gamma_{n_\mathrm{x}}$ in \eqref{eq:last:eq}  is rational with respect to $\{\phi_i\}_{i=1}^{\tau}$. 
Let $(\{r_i\}_{i=1}^{n_\mathrm{x}},s)$ be  the factorization of $\Gamma_{n_\mathrm{x}}$ 
with respect to the first $n_\mathrm{x}$ variables. Define the functions 
$\{\tilde{\alpha_i}: \Y^{n_{\mathrm x}}_0  \rightarrow \mathbb{R}\}_{i=0}^{n_{\mathrm x}-1}$ as $ \tilde{\alpha}_{i}=r_{i+1}$.
Define $\Psi:\Y^{n_{\mathrm r}}_0 \times \X_0 \rightarrow \Y^{n_{\mathrm x}}_0$ as follows
\begin{equation}\label{eq:psi:calc} \Psi(\mu,x)=\left[\begin{array}{cccc} \mu^\top &  L_f^{n_\mathrm{r}}h(x) & \cdots & L_f^{n_\mathrm{x}-1}h(x) \end{array}\right]^\top, \end{equation} for all 
$\mu \in \Y_{0}^{n_{\mathrm r}}$, $x \in \X_{0}$. Notice that  $\zeta=\Psi(\mu,x)$, if $(\mu, x) \in {\mathbb{V}}= \mathbb{Y}_0^{n_\mathrm{r}}\times \mathbb{X}_0$.  
Define now $\hat{\alpha}_i: \mathbb{V} \rightarrow \R$ and $\hat{\beta}_i: \mathbb{V} \rightarrow \R$ by
\[ 
     \hat{\alpha}_i(\mu,{x})=\tilde{\alpha}_i(\Psi(\mu,{x})), \quad
     \hat{\beta}_i(\mu,{x})=\tilde{\beta}_i(\Psi(\mu,{x}))
\]
for all $\mu \in \Y_0^{n_{\mathrm r}}$, ${x} \in \X_0$. %{\color{red}
%Let $\EuScript{T}_1 \in \mathbb{R}^{n_y \times n_y}$ such that for any $x \in \mathbb{X}_0, \EuScript{T}_1$ is a selection matrix w.r.t. the essential support of the functions $ Y_o^{n_r} \ni \mu \mapsto \hat{\alpha}_i(\mu,x), Y_o^{n_r} \ni \mu \mapsto \hat{\alpha}_i(\mu,x)\}_{i=1}^{n}$ undex $Y_0$. 
%Similarly, let $T_1 \in \mathbb{R}^{m_1 \times n_x}, m_2 \le n_x$ be a selection matrix such that for any $\mu \in Y_0^{n_r}$, $\EuScript{T}_2$ is a selection matrix of $\{ \mathbb{X}_0 \ni x \mapsto \hat{\alpha}_i(\mu,x),\mathbb{X}_0 \ni x \mapsto \hat{\beta}_i(\mu,x)\}_{i=1}^{n}$ undex $X_0$.}
Let $\EuScript{T}_1\in\mathbb{R}$  such that for any $x \in \mathbb{X}_0$, $\EuScript{T}_1$ is the selection matrix of the essential support of the functions $\{\mu \mapsto \hat{\alpha_i}(\mu,x), \mu \mapsto \hat{\beta_i}(\mu,x)\}_{i=0}^{n_{\mathrm x}-1}$ under $ \Y_0$ (essential support w.r.t. the the variables $\{\mu_i\}_{i=1}^{n_{\mathrm r}}$).
Similarly, let $\EuScript{T}_2\in\mathbb{R}^{m_2 \times n_\mathrm{x}}$  with $m_2 \le n_{\mathrm x}$ be the selection matrix of the essential support of the functions $\{\hat{\alpha_i},\hat{\beta_i}\}_{i=0}^{n_{\mathrm x}-1}$ under $ \X_0$ w.r.t.\ $x$. If $\EuScript{T}_1$ is zero, then let $\EuScript{T}=[\ 0\ \EuScript{T}_2\ ]$ and $\EuScript{P}=[\ 0_{m_2\times n_\mathrm{r}}\ \EuScript{T}_2\ ]$; otherwise, let $\EuScript{T}=\mathrm{diag}(1,\EuScript{T}_2)$ and $\EuScript{P}=\mathrm{diag}(I_{n_\mathrm{r}\times n_\mathrm{r}},\EuScript{T}_2)$.  
%
%Let $\MOD{\EuScript{P}}$ be the selection matrix of the essential support of the functions $\{\hat{\alpha_i},\hat{\beta_i}\}_{i=0}^{n_{\mathrm x}-1}$  
%\MOD{under $\mathbb{V}$}. \MOD{Partition $\EuScript{P}$ to} $\EuScript{T}_1\in\mathbb{R}^{m_1 \times n_\mathrm{r}}$ and $\EuScript{T}_2 \in \mathbb{R}^{m_2 \times n_{\mathrm x}}$ with $m_1 \le \MOD{n_{\mathrm r}}$ and $m_2 \le \MOD{n_{\mathrm x}}$ satisfying 
%\begin{equation}
%\label{thm:2:ess:support1} 
% \MOD{ \EuScript{P}}\begin{bmatrix} \mu_1 & \cdots & \mu_{n_{\mathrm r}} & \MOD{x}^\top \end{bmatrix}^{\top}=\begin{bmatrix} \begin{bmatrix} \mu_ 1&  \ldots & \mu_{n_{\mathrm r}} \end{bmatrix}\EuScript{T}_1^\top\  &\ \MOD{x}^\top \EuScript{T}_2^\top  \end{bmatrix}^{\top}.
%\end{equation}
%If \MOD{$\EuScript{T}_1$ is a zero} matrix, let $\EuScript{T}=[\ 0\ \EuScript{T}_2\ ]$ otherwise, let $\EuScript{T}=\mathrm{diag}(1,\EuScript{T}_2)$. 
Using the notation and assumptions above, we can now state the following theorem. 
\begin{theorem}[LPV embedding, full observability form] \label{thm:2}
Under the conditions of Theorem \ref{thm:1}, if $s=0$, \emph{i.e.}, factorization of $\Gamma_{n_\mathrm{x}}$ is possible without a remainder, then the LPV-SS representation \eqref{e:lpv_obvs_n}
with coefficient functions $\{\alpha_i:=\hat{\alpha}_i \circ \EuScript{P}^{-1},\ \beta_i:=\hat{\beta}_i \circ \EuScript{P}^{-1} \}_{i=0}^{n_{\mathrm x}-1}$
and with 
\begin{equation}
p=\EuScript{T} [\begin{array}{cc}y^\top & x^\top\end{array}]^\top  
 \end{equation}
and $\mathbb{P}=\EuScript{T}(\Y_0 \times \X_0)$ satisfies \eqref{eq:behav:SS1}
where
 \begin{equation}
\pi_p \mathfrak{B}_\mathrm{SS}^\mathrm{o}=\bigl\{ (u,x,y)\in  \mathcal{C}_{n_\mathrm{x}}(\R,\U_0\times\mathbb{X}_0 \times \Y_0) \mid 
\exists p \in  \mathcal{C}_{n_\mathrm{r}}(\R,\mathbb{P}), 
\exists z \in  \mathcal{C}_{n_\mathrm{x}}(\R,\Y_0^{n_{\mathrm x}}) \text{ s.t.\ (\ref{eq:ch3:03}--b) hold, while } x=\Phi^\dag(z)  \bigr\}. 
\end{equation}
\end{theorem}
\begin{proof}The proof of Theorem \ref{thm:2} follows the same line of reasoning as Theorem \ref{thm:1} and hence it is skipped. Note that the construction of $\Phi$ implies that $\{\alpha_i \diamond p,\beta_i \diamond p\}_{i=0}^{n_{\mathrm x}-1}$ will not
 depend on the derivative of $x$.
 \end{proof}
In contrast with the procedure in Section~\ref{s:nl2lpvobvs_n}, $p$ here does not include $u$; however, part of it depends on the availability of the original states of the NL system.

%<<<<<<<<<<<<<<<<<<<<<<<<<<<<<<<<<<<<>>>>>>>>>>>>>>>>>>>>>>>>>>>>>>>>>>>>>>>>>>
\section{Numerical examples}\label{s:ex}
%<<<<<<<<<<<<<<<<<<<<<<<<<<<<<<<<<<<<>>>>>>>>>>>>>>>>>>>>>>>>>>>>>>>>>>>>>>>>>>

First, two academic examples are presented to illustrate the properties of the conversion procedures discussed in the previous sections. In the first one, the relative degree is equal to the order of the system while in the second one, it is less. These examples are followed by the examples of a magnetic levitation system and an unbalanced disc system. In the latter two examples, an NL model derived from first principle laws is converted  into an LPV-SS representation. In the last example, we also show empirical validation of the model conversion both in terms of comparing responses of the real system with its LPV model and also how an LPV controller designed based on the converted model performs. 
%-------------------------------------------------------------------------------

\subsection{Conversion under full relative degree}\label{ss:ex1}
%-------------------------------------------------------------------------------
Consider a SISO NL system model \eqref{e:nl} with $n_\mathrm{x}=3$ and
\begin{equation*}
   f(x) = \bbma
0 \\
    x_1+x_3^2 \\
    x_2+x_2x_3
    \ebma, \quad
g(x) = \bbma
x_2^2+x^2_3 +1\\
   0 \\
   0
   \ebma, \quad
   h(x) = x_3.
\end{equation*}
As commonly done in practice, one could pick $x_3$ and $x_2$ as scheduling variables for LPV conversion to the form of \eqref{eq:ch3:04all}; however, that would require \MOD{accurate} measurements or estimates of these state variables if an LPV controller was to be designed and implemented based on such a converted model. 
Another problem would be the validity of this LPV conversion in terms of the represented solutions of the original NL model: it would not be clear under which condition the obtained LPV model is a valid representation of the NL model. 
So, let us see what the proposed method in this paper results in.
For this system, we have
\begin{align*}
   L_gh(x)    &=0,  \qquad L_gL_fh(x) =0,  \\
   L_gL^2_fh(x)&=(x_2^2 + x^2_3+1)(x_3 + 1),
\end{align*}
which gives that the relative degree is $n_\mathrm{r}=3=n_\mathrm{x}$ at each $x_0$ not belonging to the hyperplane  $ \X^{\dag}_0=\{x \in\mathbb{R}^3\mid (x_2^2 + x^2_3+1)(x_3 + 1) = 0\}$. Select $x_0=\mas{ccc}{0&0&0}^\top$ and $\mathbb{X}_\mathrm{r}$ to be any open subset of $\mathbb{R}^3 \setminus  \X^{\dag}_0$.  For the sake of simplicity, take $\mathbb{X}_\mathrm{r}=(-1,1)^3$. 
Computing \eqref{eq:ss:map} gives $z=\Phi(x)$ where
\begin{align*}
    \Phi(x) = \bbma x_3 & x_2 + x_2x_3 &  (x_3 + 1)(x_2^2 + x_3^2 + x_1)  \ebma^\top.
\end{align*}
The Jacobian of $\Phi$ is non-singular on $\mathbb{X}_\mathrm{r}$, in fact 
$\Phi$ is an analytic diffeomorphism on $\mathbb{X}_\mathrm{r}$ and its inverse is given by 
\begin{align*}
    \Phi^{\dag}(\mu)= 
   \bbma \frac{\left(\mu_3 - (\mu_1 + 1)^3\left(\mu_2^2 + \mu_1^2(\mu_1 + 1)^2\right)\right)}{(\mu_1 + 1)} &
    \frac{\mu_2}{\mu_1 + 1} &
    \mu_1
    \ebma^\top.
\end{align*}
Let $\Y_0=(-1,1)$, which is an open subset of $\R$ and satisfies $\Y_0^3 \subseteq \Phi(\mathbb{X}_\mathrm{r})$ and set $\X_0=\Phi^{\dag}(\Y_0^3)$.
Let $\U_0$ be an arbitrary open subset of $\R$  containing $0$.
The resulting $\Gamma_{n_\mathrm{x}}$ function, see \eqref{eq:state_trafo2}, is given by
\begin{equation*}
\Gamma_{n_\mathrm{x}}(\zeta)=\tfrac{\mu_2(2\mu_1 + 3\mu_3 + 3\mu_1\mu_3 + 6\mu_1^2 + 6\mu_1^3 - 2\mu_2^2 + 2\mu_1^4)+(\mu_1+1)\left(\mu_1(\mu_1+1)^2+\mu_2^2\right)\upsilon_1}{(\mu_1 + 1)^2},
\end{equation*}
where $\zeta=\begin{bmatrix} \mu_1\! &\! \upsilon_1\! &\! \ldots\! &\! \mu_{3}\! &\! \upsilon_{3}\end{bmatrix}$.
Factorization of this rational function is implemented by applying Algorithm~\ref{alg:fact1} resulting in 
\begin{align*}
r_1(\zeta)&=0,& 
  r_2(\zeta)&= -\tfrac{-2\mu_1+6\mu_1^2-6\mu_1^3-2\mu_2^2+2\mu_1^4}{(\mu_1 + 1)^2},\\
  r_3(\zeta) &= \tfrac{3\mu_2}{\mu_1 + 1},&
 r_4(\zeta) &=(\mu_1+1)(\mu_1+\tfrac{\mu_2^2}{(\mu_1+1)^2}),
\end{align*}
with  $s=0$.
Hence, 
\[\EuScript{R}^{-1}=\begin{bmatrix} 1 & 0 & 0 & 0 & 0 & 0 \\ 0 & 0 & 1 & 0 & 0 & 0 \\ 0 & 0 & 0 & 0 & 1 & 0 \\ 0 & 1 & 0 & 0 & 0 & 0 \\ 0 & 0 & 0 & 1 & 0 & 0 \\ 0 & 0 & 0 & 0 & 0 & 1 \end{bmatrix},\quad  \EuScript{T}=\begin{bmatrix} 1 & 0 \end{bmatrix}, \quad \EuScript{T}^{-1}=\begin{bmatrix} 1\\ 0 \end{bmatrix}, \]
  
Then, $\{\alpha_i=r_{i+1}\ \circ\ \EuScript{R}^{-1}\ \circ\ \EuScript{T}^{-1}\}_{i=0}^{2}$, and 
$\beta_0=r_4\ \circ\ \EuScript{R}^{-1}\ \circ\ \EuScript{T}^{-1}$ are defined on $\Y_0^2$ and with the resulting $p=y=\EuScript{T}\begin{bmatrix} y & u\end{bmatrix}^\top $:
\begin{align*}
\alpha_0\diamond p &=0, &
 \alpha_1 \diamond p &= -\tfrac{-2p+6p^2-6p^3+2p^4-2\dot{p}^2}{(p + 1)^2},\\
 \alpha_2\diamond p &= \tfrac{3\dot{p}}{p + 1}, &
 \beta_0\diamond p &=(p+1)(p+\tfrac{\dot{p}^2}{(p+1)^2}).
\end{align*}
The scheduling region is $\mathbb{P}=\EuScript{T}(\Y_0 \times \U_0) = \Y_0 = (-1,1)$.
The selection of the scheduling signal $p= y$, leads to the converted LPV model \eqref{e:lpv_obvs_n} which achieves embedding of the NL behavior into the solution set of the LPV-SS representation according to Theorem \ref{thm:1}. 
It is worth to mention that for this system with $p=y$, the converted matrices have only first order dynamic dependence (dependence on $p$ and $\dot p$ only). %This has been ensured by Algorithm~\ref{alg:fact1} in the factorization step in order to avoid high order derivatives of the output which in a practical implementation would be difficult to compute. 
As a further simplification, in line with Section \ref{sec:schechoice}, one can introduce $p=\eta \diamond y=  \begin{bmatrix} y & \dot{y} \end{bmatrix}^\top$ which results in rational static dependency of $\alpha_0, \alpha_1, \alpha_2, \beta_0$ by increasing the dimension of $p$, while taking $p=\begin{bmatrix} r_2 \diamond y & r_3 \diamond y & r_4 \diamond y \end{bmatrix}^\top$  results in an affine, but conservative embedding with $\alpha_1=p_1$, $\alpha_2=p_2$, $\beta_0=p_3$.
%-------------------------------------------------------------------------------
\subsection{Conversion under low relative degree}\label{ss:ex2}
%-------------------------------------------------------------------------------

To demonstrate the properties of  the procedures presented in Section~\ref{s:nl2lpvobvs_n} and \ref{s:nl2lpvobvs}, \eqref{e:nl} is considered with  $n_\mathrm{x}=3$ and
\begin{equation*} 
   f(x) = \bbma
x_2-2x_2x_3+x_3^2 \\
    x_3 \\
    \sin(x_1)
    \ebma, \ \
g(x) = \bbma
4x_2x_3 \\
   -2x_3 \\
   0
   \ebma , \ \
   h(x) = x_3.
\end{equation*} 
The system has a relative degree $n_\mathrm{r}=2<n_\mathrm{x}$ at each $x_0$ not belonging to the hyper-surface $ \X^{\dag}_0=\{x\in\mathbb{R}^3\mid \cos(x_1)x_2x_3= 0\}$. Select $x_0=\mas{ccc}{0&0&0}^\top$ and 
let $\mathbb{X}_\mathrm{r}=(0, \frac{\pi}{2}) \times (-1,1) \times (-0.5,0.5)$.
It is clear that $\mathbb{X}_\mathrm{r}$ is an open subset of $\mathbb{R}^3 \setminus  \X^{\dag}_0$ containing $0$. 
 First consider the 
approach discussed in Section~\ref{s:nl2lpvobvs} to convert the NL representation  to the full  observability canonical form \eqref{e:lpv_obvs_n}. According to \eqref{eq:ss:map}
\[
z=\Phi(x) =\bbma
 x_3 &
\sin(x_1) &
 \cos(x_1)(x_3^2 - 2x_2x_3 + x_2)
\ebma^\top.\]
The Jacobian of $\Phi$ is non-singular on $\mathbb{X}_\mathrm{r}$; in fact, $\Phi$ is an analytic diffeomorphism  on $\mathbb{X}_\mathrm{r}$
and the inverse map is
\[
\Phi^{\dag}(\zeta)= \bbma
\sin^{-1}(\zeta_2) &
\frac{-\zeta_3 + \zeta_1^2\sqrt{1 - \zeta_2^2}}{(2\zeta_1 - 1)\sqrt{1 - \zeta_2^2}} &
\zeta_1
\ebma^\top.
\]
Let $\Y_0^3 \subseteq \Phi(\mathbb{X}_\mathrm{r})$ be an open set and  $\X_0=\Phi^{\dag}(\Y_0^3)$. 
The resulting $\Gamma_{n_\mathrm{x}}  (\zeta) $ via \eqref{eq:last:eq} is given by 
\begin{equation} \label{eq:f1:eq}
\Gamma_{n_\mathrm{x}}  (\zeta) = \tfrac{- 2\zeta_2\zeta_3 - \zeta_2\zeta_3^2 + 2\zeta_2^3\zeta_3  + 2\zeta_1\zeta_2\zeta_3^2 + \left(4\zeta_1^3- 4\zeta_1^2+ \zeta_1+ 2\zeta_1\zeta_2- 2\zeta_1^2\zeta_2\right)\sqrt{(1 - \zeta_2^2)^3}}{(2\zeta_1 - 1)(\zeta_2^2 - 1)},
\end{equation}
while for all $\zeta \in \Y_0^3$,
\[
  \begin{split}
   & \tilde{\beta}_0(\zeta) =\tfrac{-2\zeta_1\left(2\zeta_2\zeta_3^2- 2\zeta_1^2\zeta_2\zeta_3\sqrt{1 - \zeta_2^2}(4\zeta_1^2- 4\zeta_1+1)\sqrt{(1 - \zeta_2^2)^3}\right)}{(2\zeta_1 - 1)(\zeta_2^2 - 1)} \\
   & \tilde{\beta}_1(\zeta) =\tfrac{-4\zeta_1\left(\zeta_3 - \zeta_1^2\sqrt{1 - \zeta_2^2}\right)}{2\zeta_1 - 1}, \quad \tilde{\beta}_2(\zeta)=0. \\
  \end{split}
\]
Finally, the factorization step is performed for the function $\Gamma_{n_\mathrm{x}}  (\zeta) $ via Algorithm~\ref{alg:fact1} as $\Gamma_{n_{\mathrm x}}$ is rational in the considered sense with  $\phi_1(\zeta)=\sqrt{1-\zeta_2^2}$, which yields the following functions
\begin{align*}
r_1(\zeta)=\tilde{\alpha}_0(\zeta)&= 
\tfrac{(4\zeta_1^2-4\zeta_1+1)\sqrt{1-\zeta_2^2}}{(2\zeta_1 - 1)(\zeta_2^2 - 1)}, \\
r_2(\zeta)=\tilde{\alpha}_1(\zeta)&= 
\tfrac{(-2\zeta_1-2\zeta_1^2)\sqrt{(1-\zeta_2)^3}}{(2\zeta_1 - 1)(\zeta_2^2 - 1)}, \\
r_3(\zeta)=\tilde{\alpha}_2(\zeta)&= 
\tfrac{(-2\zeta_2-\zeta_2+2\zeta_2^3+2\zeta_1\zeta_2\zeta_3)}{(2\zeta_1 - 1)(\zeta_2^2 - 1)},
\end{align*}
with $s=0$. According to \eqref{eq:psi:calc}, computing $\zeta=\Psi(\mu,x)$ gives that $\zeta_1=\mu_1$, $\zeta_2=\mu_2$, $\zeta_3=\sqrt{1-\mu_2^2}(\mu_1^2-2x_2\mu_1+{x}_2)$ for all $(\mu,{x}) \in {\mathbb{V}}=\mathbb{Y}_0^2 \times \mathbb{X}_0$. This results in 
$$
{\EuScript{T}_1=1, \quad \EuScript{T}_2=\begin{bmatrix} 0& 1& 0 \end{bmatrix}, \quad
\EuScript{T}=\begin{bmatrix} 1 & 0 & 0 & 0 \\ 0 & 0 & 1 & 0 \end{bmatrix},  \quad \EuScript{P}=\begin{bmatrix} 1 & 0 & 0 & 0 & 0 \\ 0 & 1 & 0 & 0 &0 \\ 0 & 0 & 0& 1 & 0 \end{bmatrix}.} 
$$
yielding  $p=\begin{bmatrix}  y & x_2 \end{bmatrix}^\top=\EuScript{T}\begin{bmatrix} y & x^\top \end{bmatrix}^\top $ with $\mathbb{P}=\EuScript{T}(\mathbb{Y}_0 \times \mathbb{X}_0)$. The resulting coefficients  are 
\begin{subequations}
\label{eq:f1:eq1}
\begin{align}
 \alpha_0 \diamond p &= \tfrac{2\dot p_1\sqrt{1-\dot p_1^2}(p_1^2-2p_2p_1+p_2)^2  + \left(4p_1^2  - 4p_1 + 2\dot p_1 + 1 - 2p_1\dot p_1\right)\sqrt{(1 - (\dot p_1)^2)^3}}{(2p_1 - 1)(\dot p_1^2 - 1)},\\
 \alpha_1 \diamond p &=   \tfrac{(4p_1^2-4p_1+1)\sqrt{1-\dot p_1^2}}{(2p_1 - 1)(\dot p_1^2 - 1)}, \\
\alpha_2 \diamond p  &=  \tfrac{(-2\dot p_1-\dot p_1+2\dot p_1^3+2p_1\dot p_1\sqrt{1-\dot p_1^2}(p_1^2-2p_2p_1+p_2))}{(2p_1 - 1)(\dot p_1^2 - 1)}, \\
\beta_0 \diamond p &= \tfrac{-4( \dot p_1(p_1^3-2p_2p_1+p_2)p_2p_1) - 2 p_1 \sqrt{1 - \dot p_1^2} (1-2p_2)}{(2p_1 - 1)(\dot p_1^2 - 1)}, \\
 \beta_1 \diamond p  &=\tfrac{4p_2p_1\sqrt{1 - \dot p_1^2}}{2p_1 - 1}, \\
\beta_2 \diamond p  &= 0.
\end{align}
\end{subequations}
where $p\in \mathscr{C}_{\infty}(\R,\mathbb{P})$.

Consider the conversion procedure introduced in Section~\ref{s:nl2lpvobvs_n}. The  map $\Phi$ is determined by
\begin{align*}
\Phi(x,u)=&\bbma
 x_3 &
\sin(x_1) &
 \cos(x_1)(x_3^2 - 2x_2x_3 + x_2 + 4x_2x_3u)
\ebma^\top.
\end{align*}
Notice that
 \[ 
   \nabla \Phi |_{x=0, u=0} = \begin{bmatrix} 0 & 0 & 1 \\ 1 & 0 & 0 \\ 0 & 1 & 0 \end{bmatrix} ,
\]
is full row rank, hence, by Lemma \ref{ss:rn_1:lemma1}, there exist compact open sets $0\in \X_0 \subseteq \mathbb{X}_\mathrm{r}=(0,\frac{\pi}{2}) \times (-1,1) \times (-0.5, 0.5)$, $\Y_0 \subseteq \R$,
$0 \in \U_0 \subseteq \R$, and an analytic map $\Phi^\dag:\Y_0^3 \times \U_0 \rightarrow \X_0$, such that
${\mu}=\Phi({x},\upsilon) \iff {x}=\Phi^{\dag}({\mu},\upsilon)$ for all $\upsilon \in \U_0, {\mu} \in \Y_0^3, {x} \in \X_0$. 
In this case, 
\begin{align*}
 \Phi^\dag ({\mu},\upsilon) =& \bbma
\sin^{-1}({\mu}_2) &
\frac{-{\mu}_3 + {\mu}_1^2\sqrt{1 - {\mu}_2^2}}{( 2{\mu}_1 - 4{\mu}_1\upsilon - 1)\sqrt{1 - {\mu}_2^2}} &
{\mu}_1
\ebma^\top.
\end{align*}
According to Corollary \ref{e:nl:obs:cor}, $\Gamma_{n_\mathrm{x}}$ {with $\zeta=\begin{bmatrix} \mu_1\! &\! \upsilon_1\! &\! \ldots\! &\! \mu_{3}\! &\! \upsilon_{3}\end{bmatrix}$} is given by:
\begin{multline}\label{eq:f2:eq}
 \Gamma_{n_\mathrm{x}} ({\zeta})=\tfrac{\mu_2\mu_3^2 - 2\mu_1\mu_2\mu_3^2 - 2\mu_2\mu_3(\mu_2^2 - 1) + 4\upsilon_1\mu_1\mu_2\mu_3^2 + 4\upsilon_1\mu_2\mu_3(\mu_2^2 - 1) + 4\upsilon_2 \mu_1\mu_3(\mu_2^2 - 1)}{(\mu_2^2 - 1)(4\upsilon_1\mu_1 - 2\mu_1 + 1)} \\
+ \tfrac{(4\mu_1^2 - 4\mu_1^3 - \mu_1 + 16\upsilon^2_1\mu_1^2 - 48\upsilon^2_1\mu_1^3 + 32\upsilon^3_1\mu_1^3  + 2\upsilon_1\mu_1 - 2\mu_1\mu_2  - 16\upsilon_1\mu_1^2 + 24\upsilon_1\mu_1^3 + 4\upsilon_2 \mu_1^3 + 2\mu_1^2\mu_2   - 4\upsilon_1\mu_1^2\mu_2)\sqrt{(1 - \mu_2^2)^3}}{(\mu_2^2 - 1)(4\upsilon_1 \mu_1 - 2\mu_1 + 1)}.
\end{multline}

Then, the factorization step is performed for $\Gamma_{n_\mathrm{x}}$ with respect to the first $4$
variables. $\Gamma_{n_{\mathrm x}}$ is rational in the considered sense with  $\phi_1{(\zeta)}=\sqrt{1-\mu_2^2}$, hence the resulting factorization is
$(\{r_i\}_{i=1}^{4}, s=0)$, {where}
\begin{subequations}
\label{eq:f31:eq}
\begin{align}
    r_1({\zeta}) & = \tfrac{(- 4\mu_1-4\mu_1^2+4\upsilon_2\mu_1^2)\sqrt{(1-\mu_2^2})^3}{(\mu_2^2 - 1)(4\upsilon_1\mu_1 - 2\mu_1 + 1)}, \\
r_2({\zeta}) &= \tfrac{(-2\mu_1+2\mu_1^2)\sqrt{(1-\mu_2^2)^3}}{(\mu_2^2 - 1)(4\upsilon_1\mu_1 - 2\mu_1 + 1)},\\
    r_3({\zeta})  & = \tfrac{(4\upsilon_2\mu_1-2\mu_2)(\mu_2^2-1)-2\mu_2\mu_3-2\mu_1\mu_2\mu_3}{(\mu_2^2 - 1)(4\upsilon_1\mu_1 - 2\mu_1 + 1)}, \\   
r_4({\zeta})&=  \tfrac{4\mu_1\mu_2\mu_3^2+4\mu_1\mu_3(\mu_2^2-1)+\sqrt{(1-\mu_2^2)^3}(-4\mu_1^2\mu_2+24\mu_1^3-16\mu_1^2+2\mu_1+32\upsilon_1^2\mu_1^3-48\upsilon_1\mu_1^2-16\upsilon_1\mu_1^2)}{(\mu_2^2 - 1)(4\upsilon_1\mu_1 - 2\mu_1 + 1)},
\end{align}
\end{subequations}
which holds for all ${\zeta \in  (\Y_0\times \U_0)^3}$and  \[
   \EuScript{T}=\EuScript{T}^{-1}=I, \quad
    \EuScript{R}^{-1}=\begin{bmatrix} 1 & 0 & 0 & 0 & 0 & 0 \\ 0 & 0 & 1 & 0 & 0 & 0 \\ 0 & 0 & 0 & 0 & 1 & 0 \\ 0 & 1 & 0 & 0 & 0 & 0 \\ 0 & 0 & 0 & 1 & 0 & 0 \\ 0 & 0 & 0 & 0 & 0 & 1 \end{bmatrix}, \]
due to the full joint essential
support of $\{r_i\ \circ\ \EuScript{R}^{-1} \}_{i=1}^{n_\mathrm x}$.
The system is embedded into the LPV-SS  form \eqref{e:lpv_obvs_n}, as described in
Theorem \ref{thm:1}, with $\mathbb{P}=\Y_0 \times \U_0$ and 
$\{\alpha_i\}_{i=0}^{3}$, $\beta_0$ satisfying 
\begin{subequations}
\label{eq:f3:eq}
\begin{align}
  \alpha_0 \diamond p & = \tfrac{(- 4p_1-4p_1^2+4\dot p_2p_1^2)\sqrt{(1-\dot p_1^2})^3}{(\dot p_1^2 - 1)(4p_2p_1 - 2p_1 + 1)}, \\
 \alpha_1 \diamond p &= \tfrac{(-2p_1+2p_1^2)\sqrt{(1-\dot p_1^2)^3}}{(\dot p_1^2 - 1)(4p_2p_1 - 2p_1 + 1)},\\
  \alpha_2 \diamond p  & = \tfrac{(4\dot p_2p_1-2\dot p_1)(\dot p_1^2-1)-2\dot p_1\ddot p_1-2p_1\dot p_1\ddot p_1}{(\dot p_1^2 - 1)(4p_2p_1 - 2p_1 + 1)}, \\   
   \beta_0 \diamond p&=  \tfrac{4p_1\dot p_1\ddot p_1^2+4p_1\ddot p_1(\dot p_1^2-1)+\sqrt{(1-\dot p_1^2)^3}(-4p_1^2\dot p_1+24p_1^3-16p_1^2+2p_1+32p_2^2p_1^3-48p_2p_1^2-16p_2p_1^2)}{(\dot p_1^2 - 1)(4p_2p_1 - 2p_1 + 1)},
\end{align}
\end{subequations}
for all $p \in \mathscr{C}_{\infty}(\R,\mathbb{P})$. Note that the resulting LPV-SS model has  $2^\mathrm{nd}$-order dynamic dependency on $p_1=y$ and only static dependency on $p_2=u$. Furthermore,  $\mathbb{P}$ can be chosen to be any open subset of $\{ ({y},{u}) \in \mathbb{R} \times \mathbb{R} \mid ({y}^2 - 1)(4{uy} - 2{y} + 1)\neq 0 \}$. 

%-------------------------------------------------------------------------------
\subsection{Magnetic levitation system}\label{ss:maglev}
%-------------------------------------------------------------------------------

To show how the proposed methodology performs in practical applications, consider a magnetic levitation system, discussed in \cite{SuShIm93}, which consists of an iron ball, an electromagnet and a photo diode based position sensor. The iron ball is levitated by  the attractive force of the electromagnet, which is controlled by an applied voltage (input signal). The model of  the system can be represented in the form of \eqref{e:nl} with 
\begin{equation*}
   f(x) = \bbma
x_2\\
G - \frac{Qx_3^2}{2M(\delta + x_1)^2} \\
\frac{x_3(2\delta + x_1)(Qx_2 - R(\delta + x_1)^2)}{(\delta + x_1)\left((L + Q)(2\delta + x_1) + Q\right)}
    \ebma , \ 
g(x) = \bbma
0 \\
  0 \\
   \frac{(\delta + x_1)(2\delta + x_1)}{(L + Q)(2\delta + x_1) + Q}
   \ebma,  
\end{equation*}
and $h(x) = x_1$ corresponding to $n_\mathrm{x}=3$ together with the parameter values given in Table~\ref{tab:maglev}.  The control objective for this system is to keep  the  distance $x_1$ (the output signal) of the ball from the magnet close to some level $\delta_\mathrm{min}\leq \delta \leq\delta_\mathrm{max}$, where $\delta_\mathrm{min}>0$ corresponds to the minimal distance of the ball from the magnet, while $\delta_\mathrm{max}$ corresponds to the maximum allowed height of levitation. The system has a relative degree $n_\mathrm{r}=3=n_\mathrm{x}$  at each $x_0$ not belonging to the hyperplane $ \X^{\dag}_0=\{x\in\mathbb{R}\mid \delta+x_1= 0\}$.  Note that this is physically always satisfied as $x_1$ must be  positive otherwise the ball reaches the magnet plate.
Take $\mathbb{X}_\mathrm{r}=(\delta_\mathrm{min},\delta_\mathrm{max})^3$ and select $x_0=\mas{ccc}{\delta& 0&2\delta\sqrt{\frac{2GM}{Q}}}^\top$.
Then, \eqref{eq:ss:map} is of the form 
\[
 \Phi(x)=\bbma
x_1 &
x_2 &
 G - \frac{Qx_3^2}{2M(\delta + x_1)^2}
\ebma^\top.\]
The Jacobian of $\Phi$ is non-singular on $\mathbb{X}_\mathrm{r}$:
\[
\nabla \Phi |_{x=x_0}=\begin{bmatrix} 1 & 0 & 0 \\ 0 & 1 & 0 \\ \frac{G}{\delta} & 0 & -\sqrt{\frac{GQ}{2M\delta^2}} \end{bmatrix}.
\]
Hence, there exist open sets $x_0\in \X_0 \subseteq \mathbb{X}_\mathrm{r}$, $\Y_0 \subseteq \R$, such that $\Phi(\X_0)=\Y_0^3$ and the restriction of $\Phi$ to $\X_0$ is an analytic diffeomorphism. 
The inverse map $\Phi^{\dag}: \Y_0^3 \rightarrow \X_0$ is  
\[
\Phi^{\dagger}(\mu)= \bbma
\mu_1 &
\mu_2 &
(\delta+\mu_1)\sqrt{\frac{2M(G-\mu_3)}{Q}}
\ebma^\top,
\]
for all $\mu \in \Y_0^3$.
The resulting function $\Gamma_{n_\mathrm{x}}$ with $\zeta=\begin{bmatrix} \mu_1\! &\! \upsilon_1\! &\! \ldots\! &\! \mu_{3}\! &\! \upsilon_{3}\end{bmatrix}$, see \eqref{eq:state_trafo2}, is 
\begin{equation}
\Gamma_{n_\mathrm{x}}({\zeta})  =
\tfrac{2(G-\mu_3)\big(R(\delta+\mu_1)^2(2\delta+\mu_1)+\mu_2\left(Q+L(2\delta+\mu_1)\right)\big)}{(\delta+\mu_1)(L(2\delta+\mu_1)+Q(1+2\delta+\mu_1))}\\ -\tfrac{(2\delta+\mu_1)\sqrt{2Q(G-\mu_3)}}{\sqrt{M}(Q+(2\delta+\mu_1)(L+Q))}\upsilon_1.
\end{equation}
Then $\Gamma_{n_\mathrm{x}}$ is rational in the considered sense with $\phi_1({\zeta})=\sqrt{G-\mu_3}$
and Algorithm \ref{alg:fact1} yields the factorization $(\{r_i\}_{i=1}^{4},s)$: 
\begin{align*}
    r_1({\zeta}) &= \tfrac{2GR\mu_1^2+8GR\delta\mu_1 + 10GR\delta^2}{(\delta+\mu_1)(L(2\delta+\mu_1)+Q(1+2\delta+\mu_1))},\\[1mm]
r_2({\zeta}) & = -\tfrac{2GQ+4GL\delta+2GL\mu_1}{(\delta+\mu_1)(L(2\delta+\mu_1)+Q(1+2\delta+\mu_1))}\\[1mm]
r_3({\zeta}) &= \tfrac{-2Q\mu_2 - 2R\mu_1^3 - 4R\delta^3 - 2L\mu_1\mu_2 - 8R\delta\mu_1^2 - 10R\delta^2\mu_1 - 4L\delta\mu_2}{(\delta+\mu_1)(L(2\delta+\mu_1)+Q(1+2\delta+\mu_1))}, \\[1mm]
r_4({\zeta}) &=-\tfrac{(2\delta+z_1)\sqrt{2Q(G-\mu_3)}}{\sqrt{M}(Q+(2\delta+\mu_1)(L+Q))},
\end{align*}
with a non-factorizable term given by
\[
s({\zeta}) = \tfrac{4GR\delta^3}{(\delta+\mu_1)(L(2\delta+\mu_1)+Q(1+2\delta+\mu_1))}.
\]
Therefore, the LPV representation \eqref{e:lpv_obvs_n} for the system can be obtained, where  $p=y$ with $2^\mathrm{nd}$ order dynamic dependence (dependence on $\{\frac{d^i}{dt^i}y\}_{i=0}^{2}$) and the non-factorizable term can be handled by seeing it as a \emph{virtual input}, see Section \ref{sec:trimming}.

\begin{center}
\begin{table}[t]%
\centering
\caption{Physical parameters of the magnetic levitation system.}\label{tab:maglev}
\begin{tabular*}{350pt}{@{\extracolsep\fill}cccccc@{\extracolsep\fill}}%
\toprule
\textbf{$L$ [H]} & \textbf{$R$ [Ohm]} & \textbf{$M$ [kg]} & \textbf{$G$ [m/s$^2$]} & \textbf{$\delta$ [m]} & \textbf{$Q$ [Hm]} \\
\midrule
2.05  & 27.03  & 0.357  & 9.807 & 0.0078  & 0.0044  \\
  \bottomrule
\end{tabular*}
\end{table}
\end{center}

%-------------------------------------------------------------------------------
\subsection{Unbalanced disc system}\label{ss:UNB}
\begin{figure} \centerline{
    \includegraphics[width=2.2in]{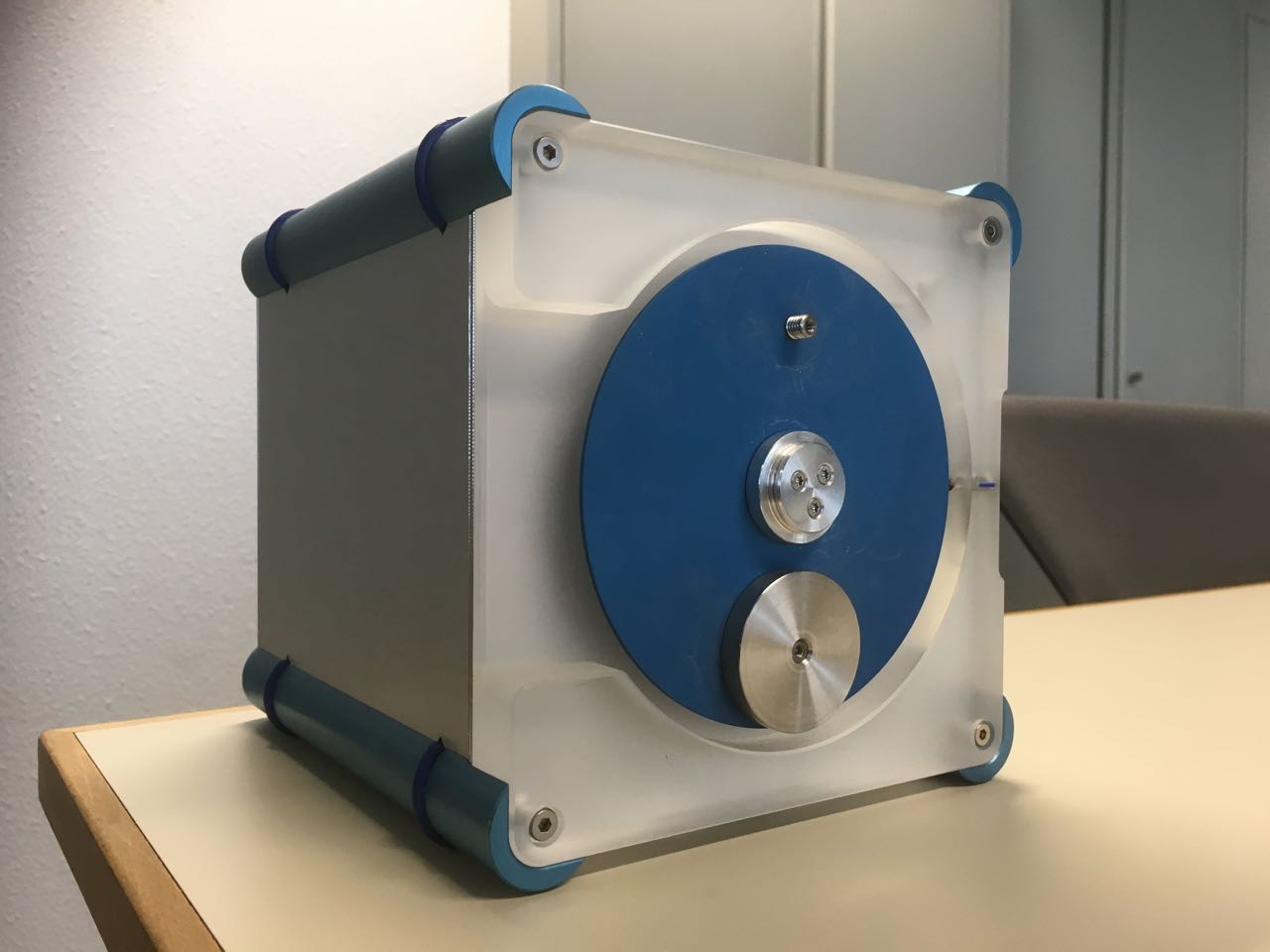} }
    \caption{Unbalanced disc system: DC motor connected to a disc with added weight. The overall system functions as a rotational pendulum.}
    \label{motor}
\end{figure}

\begin{figure}
	\centering
	\subfigure[Input signal $u$]{\includegraphics[scale=.95]{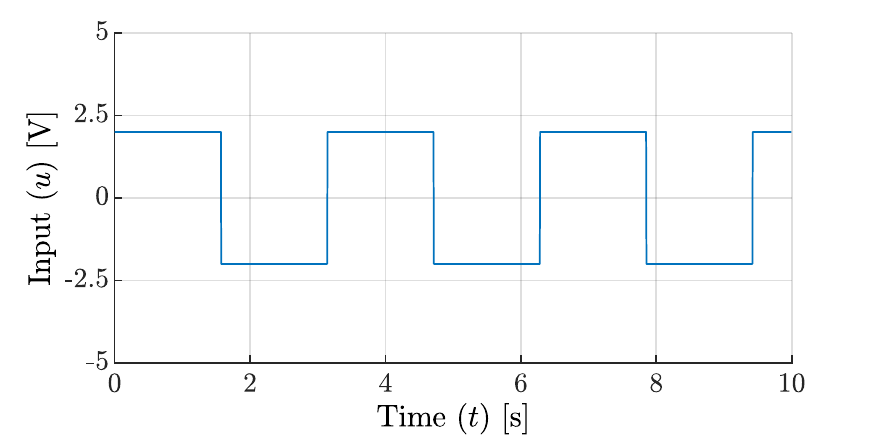}}	
	\subfigure[Output responses \legendline{mblue}measurement,\legendline{morange}nonlinear model, \legendline{myellow,dashed}LPV model]{\includegraphics[scale=0.95]{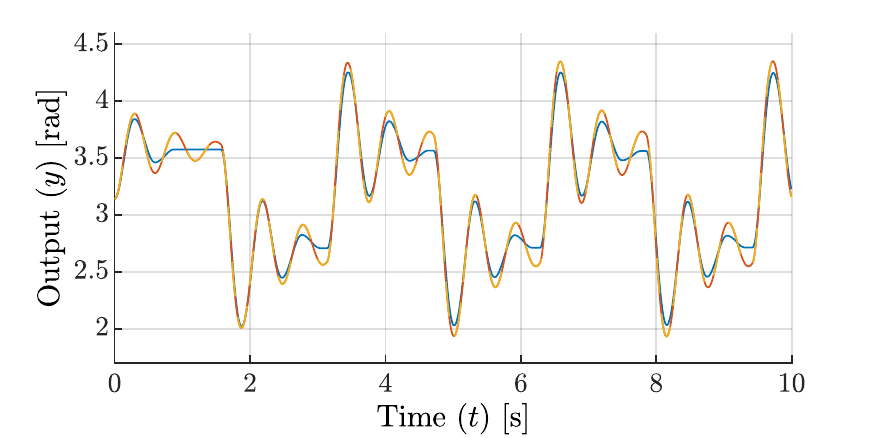}}
	\caption{Empirical validation of the identified nonlinear model and the converted LPV model.} \label{fig:resp}
\end{figure}

%-------------------------------------------------------------------------------
As \MOD{an additional} example, we demonstrate empirically  the applicability of the proposed method. Consider the unbalanced disc system depicted in Figure \ref{motor}. The dynamic behavior of this system can be well described using the following motion equations where the fast electrical subsystem is neglected
\begin{subequations}
\begin{align}\label{eq:disc}
    \dot{\theta}(t) &= \omega(t),\\
    \dot{\omega}(t) &= \tfrac{M g l}{J}\sin(\theta(t)) -\tfrac{1}{\tau}\omega(t)+\tfrac{K_m}{\tau}u(t),
\end{align}
\end{subequations}
where $\theta$ is the angular position of the mass, $\omega$ is the angular velocity of the mass and $u$ is the applied voltage on the motor. Note that $\theta$ is measurable via an encoder and it corresponds to the output of the plant. The physical parameters of  \eqref{eq:disc} have been estimated based on measurement data collected with a sampling time of $t_\mathrm{s}=0.01$sec and are given in Table \ref{table:disc}. By comparing the simulated response of the nonlinear model (using {\tt ode8} in MATLAB with fixed step-size $t_\mathrm{s}$) and the real system for a voltage signal profile that was not used in the estimation data set, we can observe from Figure \ref{fig:resp} that \eqref{eq:disc} with the estimated parameters successfully captures the physical dynamics with a 
\emph{best fit rate} (BFR)\footnote{BFR is an error measure used to compare data samples $y(k)$ ($N$ data points) w.r.t. an approximation $\hat{y}(k)$, e.g., $y$ is the measurement data and $\hat{y}$ is the response of the NL/LPV model. The BFR is computed as
\begin{equation}
	\mathrm{BFR}(y,\hat{y}) := \max\left(1-\frac{\sum_{k=1}^N(y(k)-\hat{y}(k))^2}{\sum_{k=1}^N(y(k)-\mathrm{mean}(y))^2},0\right).
\end{equation}} of $98.0$\%. Further details of the parameter estimation and the involved measurement signals can be found in \cite{TUECS2019}.

By reformulating \eqref{eq:disc} in terms of  a SISO NL state-space model \eqref{e:nl} with $x=\mas{ccc}{\theta&\omega}^\top$ and
\begin{equation*}
   f(x) = \bbma
    x_2 \\
   \tfrac{M g l}{J}\sin(x_1) -\tfrac{1}{\tau}x_2
    \ebma, \quad
g(x) = \bbma
0\\
  \tfrac{K_m}{\tau}
     \ebma, \quad
   h(x) = x_1,
\end{equation*}
we can apply the procedures presented in Section~\ref{s:nl2lpvobvs_n} to obtain an LPV model of the system.

In this case, $ L_gh(x) =0$ and $L_gL_fh(x) =\frac{K_m}{\tau}$ 
which gives that the relative degree is $n_\mathrm{r}=2=n_\mathrm{x}$ on $\mathbb{R}^2$. Select $x_0=\mas{ccc}{0&0}^\top$ and, for the sake of simplicity, $\mathbb{X}_\mathrm{r}=(-\pi,\pi)^2$. 
Computing \eqref{eq:ss:map} gives $z=\Phi(x) = \bbma x_1 & x_2   \ebma^\top$, which is an analytic diffeomorphism with $  \Phi^{\dag}(\mu)= \bbma \mu_1 & \mu_2   \ebma^\top$.
Let $\Y_0=(-\pi,\pi)$, which satisfies $\Y_0^2 =\Phi(\mathbb{X}_\mathrm{r})$ and set $\X_0=\Phi^{\dag}(\Y_0^2)=\mathbb{X}_\mathrm{r}$. Let $\U_0$ be an arbitrary open subset of $\R$  containing $0$.
The resulting $\Gamma_{n_\mathrm{x}}$ function, see \eqref{eq:state_trafo2}, is given by
\begin{equation*}
\Gamma_{n_\mathrm{x}}(\zeta)=\frac{M g l}{J}\sin(\mu_1) -\tfrac{1}{\tau}\mu_2 +\tfrac{K_m}{\tau}\upsilon_1,
\end{equation*}
where {$\zeta=\begin{bmatrix} \mu_1\! &\! \upsilon_1\! &\! \mu_{2}\! &\! \upsilon_{2}\end{bmatrix}$}. This function is polynomial with $\phi_1(\zeta)=\frac{\sin{\mu_1}}{\mu_1}=\mathrm{sinc}(\mu_1)$, and applying Algorithm~\ref{alg:fact1} results in 
\begin{align*}
r_1({\zeta})&=\tfrac{M g l}{J} \mathrm{sinc}(\mu_1),& 
  r_2({\zeta})&= -\tfrac{1}{\tau}, &
  r_3({\zeta}) &= \tfrac{K_m}{\tau},
\end{align*}
with  $s=0$.
Hence, choosing $p=\mathrm{sinc}(y)$:
\begin{align*}
\alpha_0\diamond p &=\tfrac{M g l}{J}p, &
 \alpha_1 \diamond p &= -\tfrac{1}{\tau}, & 
 \beta_0\diamond p &= \tfrac{K_m}{\tau}.
\end{align*}
The scheduling region is $\mathbb{P}=\eta(\Y_0) = \Y_0 = (-0.22,\, 1)$.
The selection of the scheduling signal $p= \mathrm{sinc}(y)$, leads to the converted LPV model \eqref{e:lpv_obvs_n} with affine static dependency that achieves embedding of the NL behavior into the solution set of the LPV-SS representation according to Theorem \ref{thm:1}.
To summarize, the NL system  \eqref{eq:disc} is embedded in the LPV representation
\begin{subequations}
\begin{align}\label{eq:discLPV}
\underbrace{\mas{c}{\dot{\theta}(t) \\ \dot{\omega}(t) }}_{\dot{x}(t)}& =\mas{cc}{0 & 1\\ \tfrac{M g l}{J}p(t)  & -\tfrac{1}{\tau} }  \underbrace{\mas{c}{{\theta}(t) \\ {\omega}(t) }}_{x(t)} +\mas{c}{0 \\ \tfrac{K_m}{\tau}}u(t), \\
y(t)&= \mas{cc}{1 & 0}x(t) ,
\end{align}
\end{subequations}
%
%
%\begin{align}
%    \dot{\theta}(t) &= \omega(t);\\
%    \dot{\omega}(t) &= \left(\tfrac{M g l}{J}p(t))\right) \theta(t) -\tfrac{1}{\tau}\omega(t)+\tfrac{K_m}{\tau}u(t);
%\end{align}
where $p(t) =  \mathrm{sinc}(\theta(t))$ with $\mathbb{P}=(-0.22,\, 1)$. By comparing the response\footnote{As the NL model is unstable, the simulated response of \eqref{eq:discLPV} is based on $p$ computed from the output of the NL simulation model.} of \eqref{eq:discLPV}, displayed in Figure \ref{fig:resp}, with the measurements and the simulated response of the NL model, it is apparent that the LPV model response is identical to the NL model simulation.

\begin{figure} \centerline{
    \includegraphics[width=3.2in]{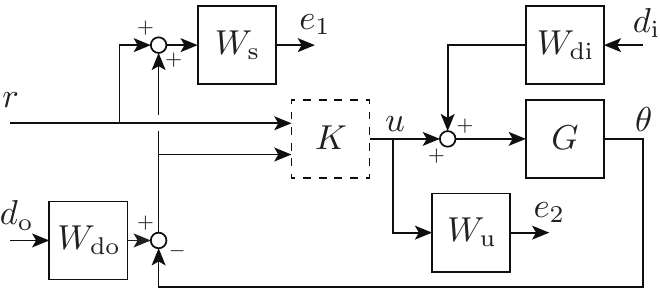} }
    \caption{Two degree of freedom control structure with mixed-sensitivity shaping for controller synthesis with the LPV model of the unbalanced disc system.} 
    \label{cont:config}
\end{figure}

\begin{figure}
	\centering
	\subfigure[Input signal $u$]{\includegraphics[scale=0.95]{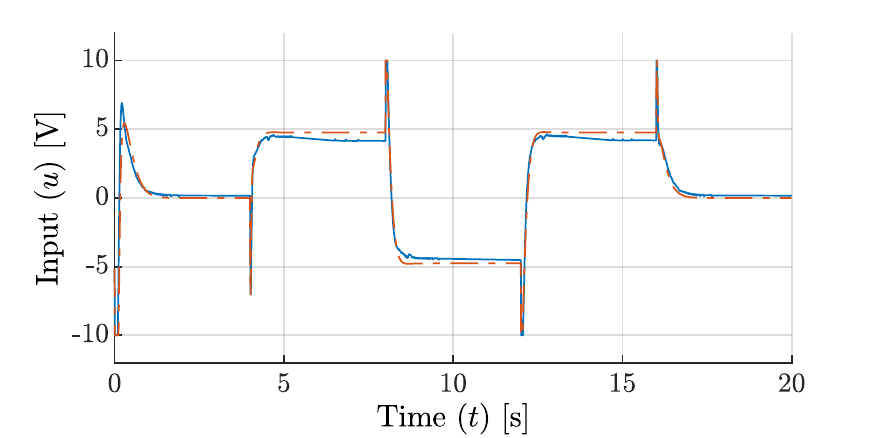}}
	\subfigure[Output signal $y$]{\includegraphics[scale=0.95]{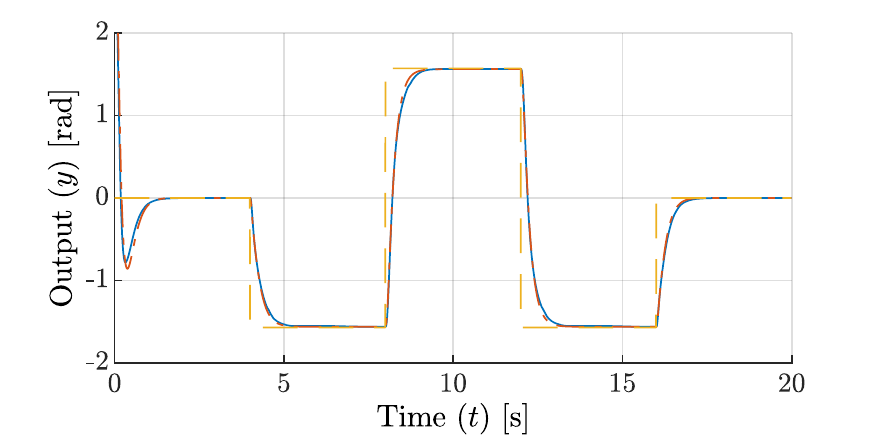}}
	\subfigure[Scheduling varaible $p$]{\includegraphics[scale=0.95]{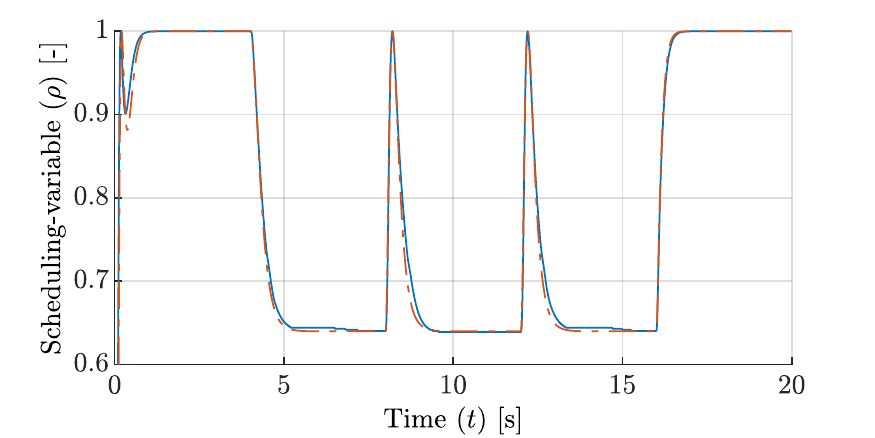}}
        \subfigure[Zoomed-in response of $y$]{\includegraphics[scale=0.95]{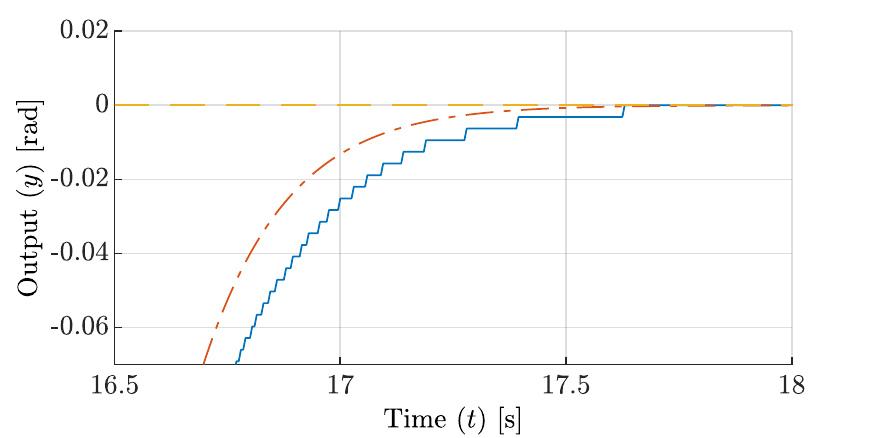}}
	\caption{Closed-loop response with the LPV controller: \legendline{mblue}experiment, \legendline{morange,dashdotted}simulation, \legendline{myellow,dashed}reference.} \label{fig:cls}
\end{figure}

A remaining question to be answered is that the resulting LPV model can be used to obtain a high-performance controller of the unbalanced disc system. For this purpose, a two degree of freedom control structure with mixed-sensitivity shaping is considered, depicted in Figure \ref{cont:config}, where $d_\mathrm{i}$ is an input disturbance, $d_\mathrm{o}$ an output disturbance and $r$ is the reference trajectory which act as disturbances to the resulting generalized plant. Furthermore, $\lbrace e_i\rbrace_{i=1}^2$ in terms of tracking error and control input are the performance channels. 
  The weighting filters are chosen as
\begin{equation}
\begin{gathered}
    W_\mathrm{s}(s) = \frac{0.5012 s + 2.005}{s+0.02005} ,\qquad W_\mathrm{u}(s) = \frac{s+40}{s+4000}, \\W_\mathrm{di} = 0.5,\qquad W_\mathrm{do} = \begin{bmatrix}0.1&0\\0&0.1\end{bmatrix}.
    \end{gathered}
    \end{equation}
Synthesis of an LPV controller by minimizing the $\mathcal{L}_2$ gain of the disturbance to performance transfer in the shaped generalized plant has been solved using polytopic synthesis based on \cite{Apkarian95TACT}. The resulting controller achieves an $\mathcal{L}_2$ bound of $0.56$, \emph{i.e.}, it successfully realizes the weighting filters encoded performance objectives. Testing the tracking capabilities of the LPV controller with the NL model \eqref{eq:disc} in simulation using a reference signal is displayed in Figure \ref{fig:cls}. The controller provides a smooth reference tracking of the NL closed-loop system with a BFR of $81.7$\%. The controller was also implemented on the real system and the measured closed-loop response is displayed in Figure \ref{fig:cls}. The achieved tracking performance\footnote{The performance increase w.r.t. to the simulation is due to the inaccuracy of the identified NL model and in other applications such inaccuracies can result in performance decrease as with any other model based approach.} in terms of BRF is $82.0$\%. This proves that the proposed LPV modeling method can be successfully applied to design an LPV controller for a nonlinear system with desired stability and performance guarantees.

%%%%%%%%%%%%
\begin{center}
\begin{table}[t]%
\centering
\caption{Identified parameters of the unbalanced disc system.}\label{table:disc}
\begin{tabular*}{350pt}{@{\extracolsep\fill}cccccc@{\extracolsep\fill}}%
\toprule
\textbf{$g$ [m/s$^2$]} & \textbf{$J$ [kg$\cdot$m$^2$]} & \textbf{$K_\mathrm{m}$ [rad/Vs$^2$]} & \textbf{$l$ [m]} & \textbf{$M$ [kg]} & \textbf{$\tau$ [1/s]} \\
\midrule
 9.8 & $2.4\cdot 10^{-4}$ & 11  & 0.041      & 0.076      & 0.40 \\
  \bottomrule
\end{tabular*}
\end{table}
\end{center}

%
%%-------------------------------------------------------------------------------
\subsection{Distillation column system}\label{ss:DC}
%%-------------------------------------------------------------------------------
\MOD{As a final example, we show how higher order derivatives of measured output signals involved in the scheduling map can be handled in the implementation of LPV controllers designed based on our LPV model conversion method. Consider the NL first principles-based model of a 4-stage binary distillation column as described in details in \cite{Sk97}. Distillation columns are commonly used in the chemical industry for component separation of liquid mixtures based on the differences in the volatility (\emph{i.e.}, boiling point) of the components. The output of the system considered here is the mole fraction of the most volatile component of the distillate product and the input is the inflow rate of the liquid to be separated.
The model is represented by \eqref{e:nl} with}
\begin{equation}\label{e:DCfunc}
\MOD{f(x) = \frac{1}{M} \bbma
 q_\mathrm{F} F(x_2 - x_1) - V\left( \delta(x_1)-x_1  \right) \\
- q_\mathrm{F} F(x_2 - x_3)+ V\left(\delta(x_1) - \delta(x_2)\right) \\
 z_\mathrm{F} F-  q_\mathrm{F} F x_3 - (1 - q_\mathrm{F})F\delta(x_3) + V\left(\delta(x_2) - \delta(x_3)\right)\\
% (1 - q_\mathrm{F})F\left(\delta(x_3)-\delta(x_4)\right) + V\left( \delta(x_3)-\delta(x_4)\right)\\
 (1 - q_\mathrm{F})F\left(\delta(x_3)-x_4\right)+V\left(\delta(x_4)-x_4\right)
\ebma, \quad
g(x) = \frac{1}{M}\bbma
x_2-x_1 \\
x_3-x_2 \\
x_4-x_3 \\
%x_5-x_4 \\
   0
\ebma, \quad
   h(x) = x_4.}
\end{equation}
\MOD{corresponding to $n_\mathrm{x}=4$ and $\delta(x_i)$ defined as
\[
\delta(x_i)=\frac{\tau x_i}{(\tau - 1)x_i + 1},\quad i=1, 2, 3, 4,
\]
where each $x_i$ stands for the mole fraction of the most volatile component (light component) in the liquid phase on tray $i$. The values of the physical/chemical parameters in \eqref{e:DCfunc} are given in Table~\ref{tab:dc} with $\tau=1.2$. The system has a relative degree $n_\mathrm{r}=2<n_\mathrm{x}$ for all $x\in\mathbb{X}^4$. Therefore, the LPV conversion can be performed by the method introduced in Section~\ref{s:nl2lpvobvs_n}. Note that the method of Section~\ref{s:nl2lpvobvs} is infeasible in a realistic application of a distillation column, as the states represent concentration levels of the liquid phase on each tray which are impossible to be accurately measured online. Hence, the procedure of Section~\ref{s:nl2lpvobvs_n} is applied. % with $x_0=\left[ \begin{array}{ccccc} 0 & 0 & 0.0881 & 0.0881 & 0.1063\end{array}\right]$. 
The map $\Phi$ of the form \eqref{eq:ss:map2}, its inverse $\Phi^\dag$ and the sets $\X_0,\U_0,\Y_0$ have been  computed according to Lemma~\ref{ss:rn_1:lemma1}, and used to compute $\Gamma_{n_{\mathrm x}}$.  The latter is used to transform the original NL model to the LPV model in the form \eqref{e:lpv_obvs_n} by factorizing the term $\Gamma_{n_\mathrm{x}}$ using Algorithm~\ref{alg:fact1}. The resulting scheduling dependence is a $3^\mathrm{rd}$-order dynamic dependence on $p=\mas{cc}{y &u}^\top$.  The exact forms of the resulting $\Gamma_{n_\mathrm{x}}$ and the factorized coefficients are not given here due to the lack of space.}
%Simulation results shown in Fig.~\ref{f:sim_DC} demonstrate that the behavior of the converted LPV model (LPV-S) is equivalent with the response of the original NL model for a typical input trajectory designed in \cite{Toth13JPC}. 
%For merely illustration purposes, the method of Section~\ref{s:nl2lpvobvs} is also applied. 
%The resulting LPV model exhibits a $2^\mathrm{nd}$-order dynamic dependence on $p_1=y$ and static dependence on $p_2=L_f^4h(x)$ and its simulated response is given in Fig.~\ref{f:sim_DC} showing equivalent behavior with the NL model up to numerical precision.

\MOD{Next, we validate the applicability of LPV control based on the obtained equivalent LPV representation when noisy output measurements are considered. To provide a realistic control scenario that respects the involved constraints of the system, we apply an LPV \emph{model predictive control} (MPC) \cite{MoNoSe20} method. MPC algorithms compute an optimal control input at each discrete time instant $k$ by solving an optimization problem based on a prediction model of the process and a cost function characterizing the performance goal (e.g., reference tracking). For this purpose, an accurate model of the process is crucial for the success of such a control methodology. The main advantage of the LPV formulation of the MPC problem is that in general it offers convex optimization based solution by trading off performance due to conservatism of the prediction model.}

\MOD{Based on the derived LPV representation of \eqref{e:DCfunc}, we can use directly the converted state in the MPC problem, which is composed of the output of the system and its derivatives up to $3^\mathrm{rd}$-order. However, the challenge here is that we need the derivatives of the output (up to order $3$), which can be obtained by numerical differentiation and hence the measurement noise can be  significantly amplified, affecting the overall performance of the closed-loop system. We also use this converted state and the input together with its derivatives up to order $2$ to compute the scheduling variable $p$, which is used to update the parameter-dependent system matrices of the prediction model at every sampling time. The exact implementation is explained later.}

\MOD{The optimization problem of the MPC considered here is formulated as follows
\begin{subequations}\label{e:MPCcost}
\begin{align}
\min_{\Delta u(0|k), \Delta u(1|k),\cdots,\Delta u(N|k)}
\sum_{i=0}^N  & (r(i|k)-y(i|k))^\top Q (r(i|k)-y(i|k)) +\Delta u^\top(i|k)R \Delta u(i|k) \\
 {\rm s.t.} \qquad \qquad  \quad & \Delta u_{\min} \leq \Delta u(i|k)\leq \Delta u_{\max},\\
&  u_{\min} \leq  u(i|k)\leq u_{\max}, \\
&  y_{\min} \leq  y(i|k)\leq y_{\max}, 
\end{align}
\end{subequations}
$i=0,1,\cdots, N$, where the argument $i|k$ indicates prediction step $i$ at instant $k$, $r$ is the reference trajectory,  $\Delta u$ represents the rate of change of $u$, $N$ is the prediction horizon and $Q\geq 0$, $R > 0$ are tuning matrices. The decision variable of the optimization problem \eqref{e:MPCcost} is $\Delta u$, and hence, we can achieve offset-free control. In order to realize such an MPC scheme, % method discussed in [WE NEED AN EXACT SPECIFICATION OF THE APPROACH WE USED]
we discretized the obtained continuous-time LPV model using the Euler's forward method, %and to achieve an offset-free control,
considered $\Delta u$ as the rate of change of the reflux, %as the decision variable (i.e. input $u$) in the MPC optimization problem. 
and as an output $y$ the purity of the top product was used. For constraints, we considered $[\Delta u_{\min}, \ \Delta u_{\max}]=[-436.25, \ +436.25]$, $[u_{\min}, \ u_{\max}]=[1175, \ 9900]$ kmol/min, and $[y_{\min}, \ y_{\max}]=[0.85, \ 0.99]$ for $\Delta u,  u, y$, respectively. %Moreover, we constrained the rate of change of $u$ to be within $\pm 436.25$.
The prediction horizon of the MPC has been taken as $N=15$, and we consider the weights of the output and the input in the MPC  cost function, which is quadratic, as $Q=10^7$ and $R=10^{-5}$, respectively. The MPC online optimization problem \eqref{e:MPCcost} is  cast as a quadratic programming problem.}

\MOD{The performance of the closed-loop system with the LPV MPC has been evaluated with $-3\%$ change in the set point of $y$, at the sampling instant $k=334$ followed by $+1.5\%$ change in the set point  at $k=668$ as shown in Fig.~\ref{fig:distcol}. At the same time, we have applied three changes of the feed flow rate $F$ as input disturbances: a $-20\%$ decrease  at $k = 167$, again a $-20\%$ decrease  at $k=501$ and a $+40\%$ increase at $k=835$. Such scenario of operation is similar to what was discussed in \cite{RaHiAzSh12}. For comparison, we carried out the simulation for two cases,  with noisy and noise-free output. In case of the noisy output, a signal-to-noise ratio of $29.5$ dB has been considered with additive white Gaussian measurement noise. To reduce  the noise effects in the numerically differentiated signals, which include $\frac{\rm d}{{\rm d}t}y$, $\frac{\rm d^2}{{\rm d}t^2}y$ and $\frac{\rm d^3}{{\rm d}t^3}y$, we used moving average filters of order $10$, $2$ and $2$, respectively. The orders were chosen to find a suitable trade-off between noise filtering, truncation of the frequency content and introduced phase lag. The output derivatives  are recursively filtered and used to construct the model represented state variables at every sample. They are used also  together with  the input and its derivatives  $\frac{\rm d}{{\rm d}t}u$ and $\frac{\rm d^2}{{\rm d}t^2}u$ to compute $p$ and hence to update the LPV model matrices during the MPC implementation. }

\begin{figure}
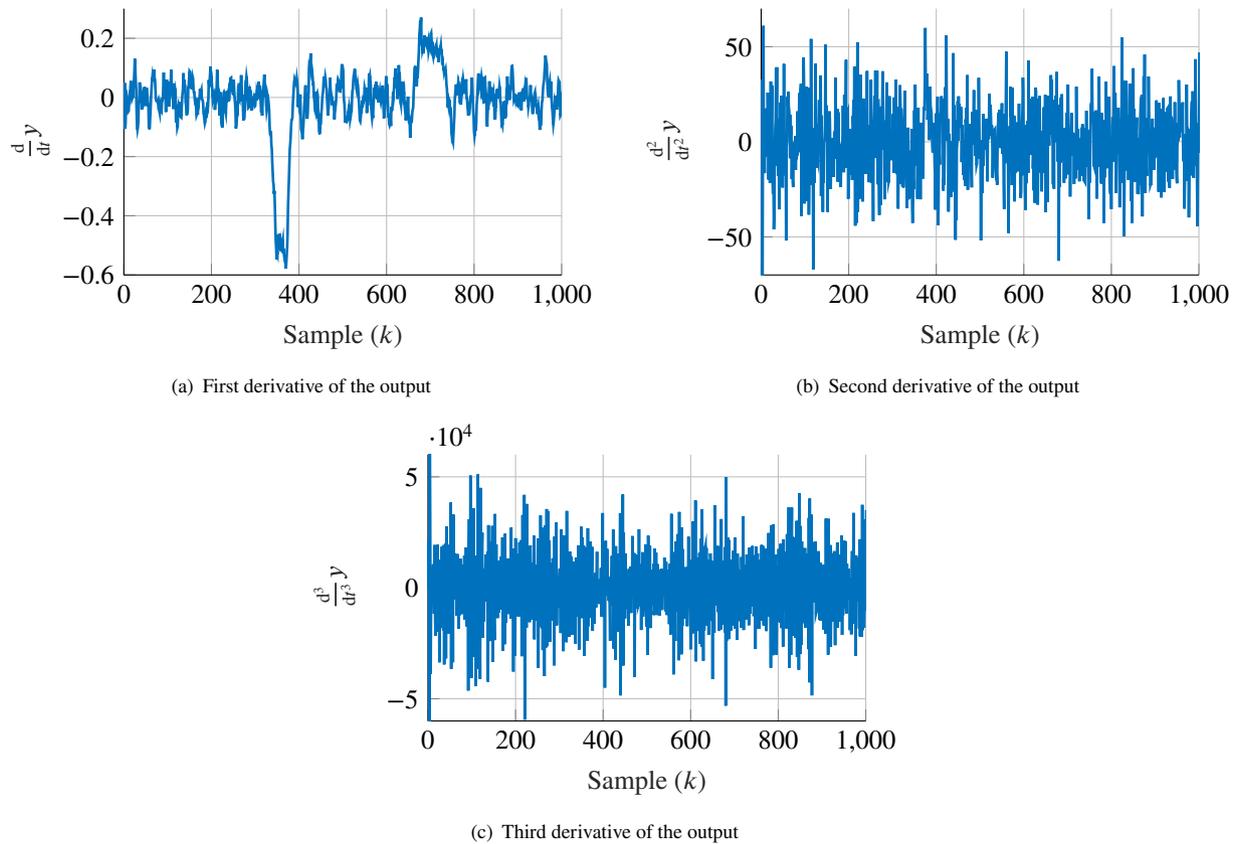

	\centering
	\subfigure[First derivative of the output]{\input{figures/ydot.tex}}\hspace{4mm}
	\subfigure[Second derivative of the output]{\input{figures/yddot.tex}}
	\subfigure[Third derivative of the output]{\input{figures/ydddot.tex}}\hspace{4mm}
	\caption{The filtered derivative of scheduling signals  used to update the  distillation column prediction model for the LPV MPC implementation.}
	 \label{fig:distcol_sch}
\end{figure}

\MOD{Based on the above discussed discrete-time implementation of the MPC controller, the closed-loop system has been simulated with the plant dynamics taken as the continuous-time NL model in  \eqref{e:DCfunc} with synchronized ZOH actuation and sampling. Figures \ref{fig:distcol}a-d show the closed-loop performance  with and without output measurement noise. Generally, the effect of the noise increases the fluctuation of the applied $u$ and slightly $y$; however, the tracking capability is still comparable to the case of noise-free $y$. In both cases, the desired set points of the output are reached within less than 50 samples with almost no overshoot and no steady-state error. The disturbance effects are successfully rejected in both cases by the MPC design. % For the input in the noisy case, the disturbances effects are not distinguishable due to its fluctuation.
The filtered derivatives of the output $y$, which are used as  scheduling signals for updating the  distillation column prediction model during the MPC implementation, are shown in Fig.~\ref{fig:distcol_sch}a-c. } %\MOD{ [WOULD BE IMPORTANT TO PLOT THE DERIVATIVES OF THE USED SCHEDULING]}

%
%\MOD{[IT WOULD BE IMPORTANT TO GIVE THE OVERALL COST AND HOW THAT CHANGES WITH CHANGING THE SNR LEVEL (TILL IT IS NOT TOO BAD)]}

\begin{center}
\begin{table}[t]%
\centering
\caption{Physical parameters of the distillation column system.}\label{tab:dc}
\begin{tabular*}{350pt}{@{\extracolsep\fill}ccccc@{\extracolsep\fill}}%
\toprule
\textbf{$M$ [kmol]} & \textbf{$z_\mathrm{F}$ [mole frac.] }  & \textbf{$F$ [kmol/min]} & \textbf{$q_\mathrm{F}$} & \textbf{$V$ [kmol/min]} \\
\midrule
 30  & 0.65 & 215 & 1.0 & 1800  \\
  \bottomrule
\end{tabular*}
\end{table}
\end{center}

%\begin{figure}[!t] \vskip -6mm\centerline{
%    \includegraphics[width=3.2in]{figures/sim.eps} }
%    \caption{Response of the distillation column system and the converted LPV models: LPV-F (LPV model in full observable form); LPV-S (LPV model in simplified observable form).} 
%    \label{f:sim_DC}
%\end{figure}

\MOD{Finally, to measure numerically the effects of the noise on the control performance, the mean square tracking errors with and without measurement noise were calculated to be $1.47\times 10^{-5}$ and $1.40\times 10^{-5}$, respectively. The quadratic cost of the MPC optimization can be seen as a performance measure, for which the average cost with and without measurement noise was  $2.37\times 10^{3}$ and $1.934\times 10^{3}$, respectively. It is larger for the noisy case by a factor of $1.22$, %[IF YOU DON'T GIVE THE EXACT VALUES IT IS IMPOSSIBLE TO UNDERSTAND WHAT 1.22 MEANS] 
which indicates that the loss of performance was not significant due to the measurement noise. Finally, we repeated the simulation with lower values of signal-to-noise ratio (SNR) but with the same tuning parameters $Q, R, N$ and filters as above and with the same seed settings for the noise generator. For an SNR of $23.5$ dB, the mean square tracking error and the average cost were 
$1.63\times 10^{-5}$ and $3.42\times 10^{3}$, respectively, which still indicate reasonable performance; however, below that value of SNR, it was necessary to tune $Q, R, N$, to avoid infeasibility of the MPC optimization problem.
}

\MOD{In summary, this example demonstrates that reasonable
closed-loop performance can be achieved with the proposed method using high-order output derivatives with noisy measurements in the scheduling map without the need of direct state measurements or nonlinear observers designed for the process.  }

\newlength\figH 
\newlength\figW
\setlength\figH{3.5cm} 
\setlength\figW{6cm} 
\begin{figure}
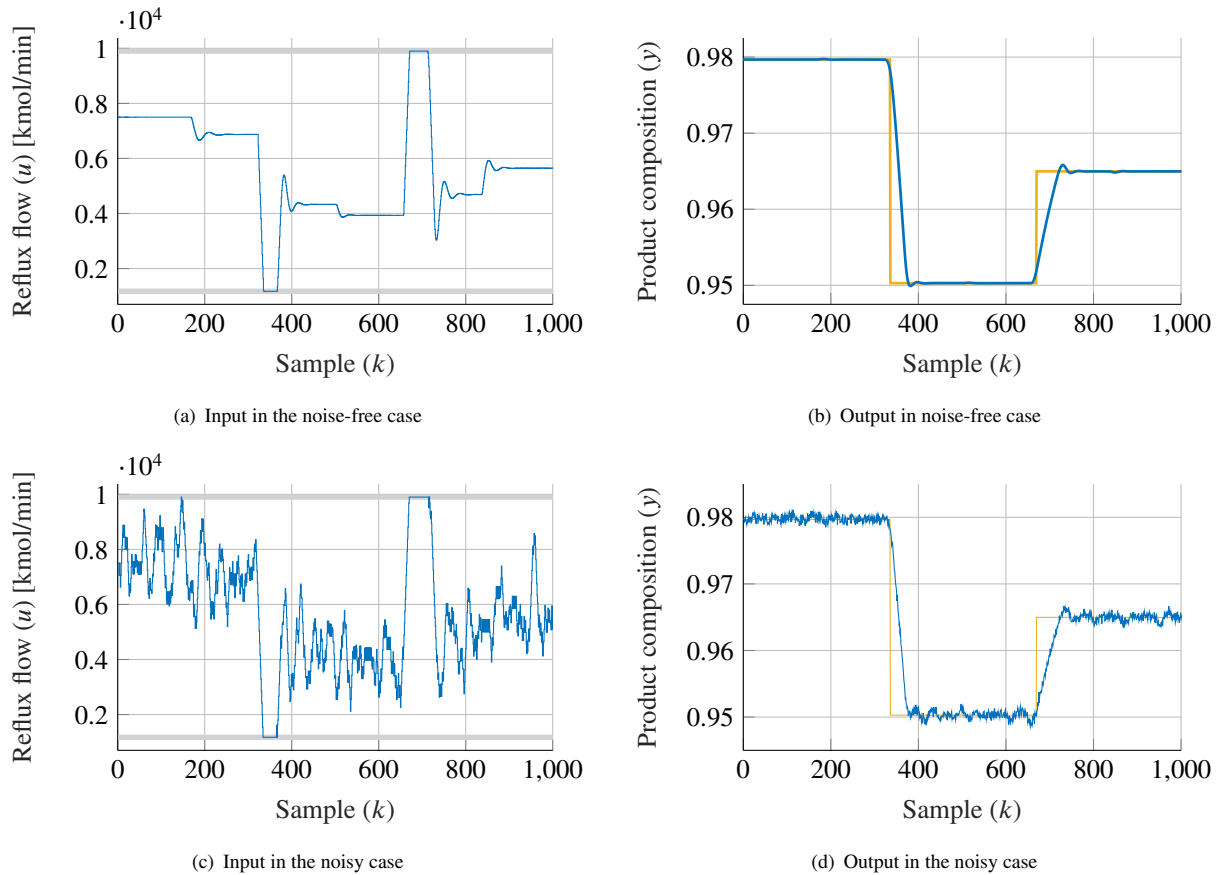

	\centering
	\subfigure[Input in the noise-free case]{\input{figures/distcol_u.tex}}\hspace{4mm}
	\subfigure[Output in noise-free case]{% This file was created by matlab2tikz.
\definecolor{mycolor1}{rgb}{0.92900,0.69400,0.12500}%
\definecolor{mycolor2}{rgb}{0.00000,0.44700,0.74100}%
\begin{tikzpicture}

\begin{axis}[%
width=0.96\figW,
height=\figH,
at={(0\figW,0\figH)},
scale only axis,
xmin=0,
xmax=1000,
xlabel style={font=\color{white!15!black}},
xlabel={Sample ($k$)},
ymin=0.9475,
ymax=0.9825,
ylabel style={font=\color{white!15!black}},
ylabel={Product composition ($y$)},
axis background/.style={fill=white},
axis x line*=bottom,
axis y line*=left,
xmajorgrids,
ymajorgrids,
legend style={legend cell align=left, align=left, draw=white!15!black}
]
\addplot [color=mycolor1, line width=1.0pt]
  table[row sep=crcr]{%
1	0.979671410229116\\
335	0.979671410229116\\
336	0.950281267922151\\
669	0.950281267922151\\
670	0.964976339075633\\
1001	0.964976339075633\\
};
%\addlegendentry{data1}

\addplot [color=mycolor2, line width=1.0pt]
  table[row sep=crcr]{%
1	0.979671410229116\\
171	0.979680081155379\\
182	0.979751386981206\\
187	0.979734604389478\\
202	0.979648034963475\\
209	0.979651370878287\\
228	0.979679857052929\\
279	0.979671771020776\\
324	0.979661584754581\\
326	0.979616958632164\\
327	0.979574427727016\\
328	0.979514622183956\\
329	0.979434478182043\\
330	0.979330612603349\\
331	0.979199201222514\\
332	0.979035847759292\\
333	0.978835436602594\\
334	0.978591962119935\\
335	0.978298327227435\\
336	0.977952045237885\\
337	0.977555963025338\\
338	0.977113631921043\\
339	0.976628775899599\\
340	0.976105128497807\\
341	0.975546321952834\\
342	0.974955814994019\\
343	0.974336848954295\\
344	0.973692424369005\\
345	0.973025292144484\\
346	0.972337954863178\\
347	0.971632674918624\\
348	0.970911487039871\\
350	0.969428480184661\\
352	0.967901259890027\\
354	0.966339545798178\\
356	0.96475094624418\\
359	0.962330235186073\\
362	0.959876808029094\\
366	0.956571675709597\\
367	0.955746206015192\\
368	0.954939014381011\\
369	0.954165112288024\\
370	0.953439121753604\\
371	0.952774999661756\\
372	0.952181627960158\\
373	0.951664395615808\\
374	0.951224952110351\\
375	0.950861208670744\\
376	0.950568648539161\\
377	0.95034100314399\\
378	0.950170962960101\\
379	0.950050859841667\\
380	0.949973102089757\\
381	0.949930476118197\\
382	0.949916328226095\\
383	0.949924646762497\\
385	0.94998793143202\\
388	0.950137784155004\\
391	0.950283739651695\\
393	0.950355518449442\\
395	0.950399920073892\\
397	0.95041707702967\\
399	0.950411192711158\\
402	0.950374058554189\\
409	0.950272125803963\\
413	0.950250166612591\\
418	0.950257410456402\\
430	0.950289578281513\\
506	0.950296060291066\\
513	0.9503405856467\\
517	0.95033486281261\\
533	0.950272243426866\\
611	0.950281270637902\\
658	0.950287975015499\\
659	0.950307719007355\\
660	0.950346763364223\\
661	0.950408976300764\\
662	0.950496545579767\\
663	0.950610272887275\\
664	0.950749881777483\\
665	0.950914306502\\
666	0.951101943984099\\
667	0.951310861987849\\
668	0.951538963448002\\
669	0.951784110776885\\
671	0.952316582300682\\
673	0.952883814798952\\
683	0.955766743510594\\
687	0.956877651301284\\
691	0.95795672699478\\
695	0.959004016639938\\
699	0.960020147435216\\
703	0.961005910720473\\
707	0.961962134308465\\
711	0.96288964323071\\
714	0.963565807595273\\
716	0.964000107118068\\
718	0.964410096467418\\
720	0.964786618211178\\
722	0.965121312840665\\
723	0.965270138512551\\
724	0.965404863693038\\
725	0.965523954838886\\
726	0.965625651992582\\
727	0.965707919487159\\
728	0.965768424889916\\
729	0.965805421580626\\
730	0.965818255417844\\
731	0.9658072469324\\
732	0.965773795761379\\
733	0.965720318885587\\
735	0.965567003287333\\
741	0.965023983710466\\
743	0.964897847123666\\
745	0.964816422986701\\
747	0.964777263914812\\
749	0.964773264397763\\
751	0.964795233227392\\
754	0.964856326096196\\
760	0.964986162822584\\
763	0.965023720653562\\
766	0.965037079269223\\
770	0.965025512418265\\
783	0.964959081396955\\
791	0.96496787557021\\
803	0.96498150000059\\
838	0.9649697089136\\
841	0.964921718127243\\
845	0.96484790288514\\
848	0.964826484651553\\
851	0.964840198901129\\
855	0.964893613913091\\
861	0.964979956589104\\
865	0.965009602014675\\
869	0.965012642679085\\
877	0.964981024944336\\
884	0.964964648155501\\
894	0.96497379951802\\
908	0.964978773236908\\
944	0.964976556922466\\
1001	0.964976337021767\\
};
%\addlegendentry{data2}

\end{axis}
\end{tikzpicture}%
}
	\subfigure[Input in the noisy case]{\input{figures/distcolnoisy_u.tex}}\hspace{4mm}
	\subfigure[Output  in the noisy case]{\input{figures/distcolnoisy_y.tex}}
	\caption{Closed-loop response of the distillation column system with the LPV MPC controller: 
	\legendline{mgray}limits \legendline{mblue}simulation, \legendline{myellow}reference.} \label{fig:distcol}
\end{figure}

\setlength\figH{3.5cm} 
\setlength\figW{6cm} 

%\begin{figure} \centerline{
%    \includegraphics[width=3.2in]{figures/distcol} }
%    \caption{} 
%    \label{f:distcol}
%\end{figure}
%
%
%\begin{figure}[h!] \centerline{
%    \includegraphics[width=3.2in]{figures/distcolnoisy} }
%    \caption{} 
%    \label{f:distcolnoisy}
%\end{figure}

%<<<<<<<<<<<<<<<<<<<<<<<<<<<<<<<<<<<<>>>>>>>>>>>>>>>>>>>>>>>>>>>>>>>>>>>>>>>>>> 

\section{Conclusions and future works}\label{s:concl} %<<<<<<<<<<<<<<<<<<<<<<<<<<<<<<<<<<<<>>>>>>>>>>>>>>>>>>>>>>>>>>>>>>>>>>>>>>>>>>
In this paper, a systematic and automated approach has been introduced to synthesize LPV  state-space representations of nonlinear systems via the idea of multi-path feedback linearization. The main advantage of the proposed approach is its ability to synthesize the model with minimal scheduling dependency where the scheduling map is based on only measurable input-output signals of the original system. This ensures implementability and minimized conservativeness of the LPV embedding. However, as demonstrated by the procedure, this often results in dynamic dependency over these signals. To avoid dynamic dependency especially over  input variables, a modified version of the approach is presented that substitutes those dependencies with  dependency relation on only part of the state variables of the original nonlinear representation.

\bibliography{ieeetr}%

%\clearpage

%\section*{Author Biography}
%
%\begin{biography}{\includegraphics[width=66pt,height=86pt,draft]{empty}}{\textbf{Author Name.} This is sample author biography text this is sample author biography text this is sample author biography text this is sample author biography text this is sample author biography text this is sample author biography text this is sample author biography text this is sample author biography text this is sample author biography text this is sample author biography text this is sample author biography text this is sample author biography text this is sample author biography text this is sample author biography text this is sample author biography text this is sample author biography text this is sample author biography text this is sample author biography text this is sample author biography text this is sample author biography text this is sample author biography text.}
%\end{biography}

\end{document}